\documentclass[10pt]{article}
\usepackage{hyperref}
\usepackage{amsthm,amsmath,amsthm,amssymb,bbm,amsfonts,dsfont}
\usepackage{graphicx}
\usepackage{subfigure}
\usepackage{color}
\usepackage{cite}
\usepackage[normalem]{ulem}

\usepackage[labelfont=bf,labelsep=period,justification=raggedright]{caption}

\makeatletter
\renewcommand{\@biblabel}[1]{\quad#1.}
\makeatother

\date{}

\def \eps {\epsilon}

\def \R {\mathbb{R}}
\def \E {\mathbb{E}}

\newcommand{\Prob}{\mathbb{P}}

\def \eqdef {\stackrel{def}{=}}

\graphicspath{{./Figures/}}

\def \E {\mathbb{E}}
\newcommand{\Exp}[1]{\mathbb{E}\left [ #1 \right]}

\theoremstyle{plain}

\theoremstyle{definition}

\begin{document}
	\begin{flushleft}
	{\Large
	\textbf{Macroscopic equations governing noisy spiking neuronal populations}
	}
	\\
	Mathieu Galtier$^{1,\ast}$, 
	Jonathan Touboul$^{2,3,\star}$, 
	\\
	\bf{1} Jacobs University, Bremen, Germany
	\\
	\bf{2} The Mathematical Neuroscience Laboratory, CIRB/ Coll\`ege de France\footnote{CNRS UMR 7241, INSERM U1050, UPMC ED 158, MEMOLIFE PSL}, Paris, France
	\\
	\bf{3} BANG Laboratory, INRIA Rocquencourt, Paris, France
	\\
	$\ast$ E-mail: m.galtier@jacobs-university.de\\
	$\star$ E-mail: jonathan.touboul@college-de-france.fr\\
	\end{flushleft}
%
%
%

\section*{Abstract}
At functional scales, cortical behavior results from the complex interplay of a large number of excitable cells operating in noisy environments. Such systems resist to mathematical analysis, and computational neurosciences have largely relied on heuristic partial (and partially justified) macroscopic models, which successfully reproduced a number of relevant phenomena. The relationship between these macroscopic models and the spiking noisy dynamics of the underlying cells has since then been a great endeavor. Based on recent mean-field reductions for such spiking neurons, we present here {a principled reduction of large biologically plausible neuronal networks to firing-rate models, providing a rigorous} relationship between the macroscopic activity of populations of spiking neurons and popular macroscopic models, under a few assumptions (mainly linearity of the synapses). {The reduced model we derive consists of simple, low-dimensional ordinary differential equations with} parameters and {nonlinearities derived from} the underlying properties of the cells, and in particular the noise level. {These simple reduced models are shown to reproduce accurately the dynamics of large networks in numerical simulations}. Appropriate parameters and functions are made available {online} for different models of neurons: McKean, Fitzhugh-Nagumo and Hodgkin-Huxley models. 
	
\section*{Author Summary}
Biologically plausible large-scale neuronal networks are extraordinary complex: they involve {an} extremely large numbers of neurons (on the order of hundreds of thousands), each of these having a complex, excitable nonlinear dynamics and subject to noise. This complexity motivated the use of deterministic low-dimensional descriptions of populations of neurons based on heuristic descriptions of the cortical function. These reduced models were successful in accounting for a number of phenomena in the brain such as spatio-temporal pattern formation. The relationship between well-accepted neuronal networks and heuristic macroscopic descriptions has since then been an important endeavor in the computational neuroscience community. We address here the problem of deriving such simple macroscopic descriptions rigorously based on the dynamics of large-scale neural networks. We apply our reduction method to the most prominent neuron models, and show that indeed, reduced models precisely account for macroscopic 
networks activity. By doing so, we can precisely compute the parameters involved in macroscopic models, and these are made available online for those neuroscientists aiming at precisely simulating large-scale networks of McKean, Fitzhugh-Nagumo or Hodgkin-Huxley neurons. Analytical developments are also provided in a simplified model to illustrate the reduction method. 

\section*{Introduction}

The activity of the brain is characterized by large-scale macroscopic states resulting of the structured interaction of a very large number of neurons. These macroscopic states correspond to the signal experimentally measured through usual recording techniques such as extracellular electrodes, optical imaging, electro- or magneto- encephalography and magnetic resonance imaging. All these experimental imaging protocols indeed record the activity of large scale neuronal areas involving of the order of thousands to millions of cells. At the cellular level, neurons composing these columns manifest highly complex, excitable behaviors characterized by the intense presence of noise. Several relevant brain states and functions rely on the coordinated behaviors of large neural assemblies, and resulting collective phenomena recently raised the interest of physiologists and computational neuroscientists, among which we shall cite the rapid complex answers to specific stimuli~\cite{thorpe-delorme-etal:01}, decorrelated activity~\cite{ecker-berens-etal:10,renart-de-la-rocha-etal:10}, large scale oscillations~\cite{buszaki:06}, synchronization~\cite{izhikevich-polychronization:06}, and spatio-temporal pattern formation~\cite{ermentrout-cowan:80,coombes-owen:05}.

All these evidences motivate the development of models of the collective dynamics of neuronal populations, that are simple enough to be mathematically analyzed or efficiently simulated. A particularly important problem would be to derive tractable macroscopic limits of the widely accepted and accurate Hodgkin-Huxley model~\cite{hodgkin-huxley:52}. However, describing the activity of a network at the cellular scale yields extremely complex, very high dimensional equations that are mathematically intractable and lead to excessively complex and time consuming numerical simulations. 

The question of the macroscopic modeling of cortical activity and their relationship with microscopic (cellular) behavior has been the subject of extensive work. Most studies rely on heuristic models (or firing-rate models) since the seminal works of Wilson, Cowan and Amari \cite{amari:72,wilson-cowan:73}. These models describe a macroscopic variable, the population-averaged firing-rate, through deterministic integro-differential or ordinary differential equations. Analytical and numerical explorations characterized successfully a number of phenomena, among which spatio-temporal pattern formation and visual illusions (see~\cite{bressloff2012spatiotemporal} for a recent review). This approach was complemented by a number of computational studies introducing noise at the level of microscopic equations, the effect of which vanishes in the limit where the number of neurons tends to infinity. These approaches are generally based on simplified neuron models and make strong assumptions on the dynamics (e.g. sparse 
connectivity~\cite{brunel-hakim:99}, Markovian modeling of the firing and van Kampen expansion~\cite{bressloff:09}). {Relationship between spiking neuronal networks and mean firing rates in simplified models and deterministic settings has also been the subject of a number of outstanding researches~\cite{rinzel-frankel:92,ermentrout:94}. These averaging techniques were based on temporal averaging of periodic spiking behaviors. For instance, in~\cite{ermentrout:94}, the author presents a reduction to Wilson-Cowan systems for the single-cell deterministic Morris-Lecar system, taking advantage of the separation of timescales between slow synapses and cell dynamics. In contrast with these researches, we propose a mixed population and temporal averaging for stochastic networks, taking advantage of the collective effects arising in large networks.}

Despite these efforts, deriving the equations of macroscopic behaviors of large neuronal networks from relevant descriptions of the dynamics of noisy neuronal networks remains today one of the main challenges in computational neurosciences, as discussed in P. Bressloff's review~\cite{bressloff2012spatiotemporal}. This is precisely the aim of the present manuscript. Inspired by statistical mechanics methods, we start from rigorously derived limits of neuronal network equations~\cite{touboul2011mean} with excitable dynamics of Hodgkin-Huxley type. The resulting equations, referred to as the \emph{mean-field equations}, are hard to interpret and to relate to physical observable quantities. In the gas dynamics domain, mean-field equations such as Boltzmann's equation were used to derive the behavior of macroscopic quantities such as the local density, macroscopic local velocity and local temperature fields, in relationship with the microscopic activity of the particles, and resulted in the derivation of the 
celebrated Navier-Stokes equations that provide important information on the fluid dynamics. In our biological case, a particularly important quantity accessible through measurement is the macroscopic variable corresponding to an averaged value, over neurons at certain spatial locations, and on a specific time interval, of the activity of each cell. By doing so, we will reduce the complex high dimensional noisy dynamics of microscopic descriptions of neural networks into a simple, deterministic equation on macroscopic observables. 

In the Material and Methods section, we will introduce the basic network equations and their mean-field limits, and describe the methodology we propose for deriving macroscopic equations. This method reduces the dynamics of the average firing-rate to the knowledge of a particular function, the effective non-linearity, which can be numerically computed in all cases. This methodology is put in good use in the Results section in the case of the McKean, Fitzhugh-Nagumo and Hodgkin-Huxley neurons. In each case, the effective-nonlinearity is numerically computed for different noise levels. The reduced low-dimensional macroscopic system is then confronted to simulations of large networks. The discussion section explores the limitations and some implications of the present approach. In the appendix, analytical developments are also provided in the case of the deterministic McKean neuron in order to illustrate the mechanics of the reduction.

\section*{Material and Methods}
In this section we introduce the original networks models, their mean-field limits and the formal derivation of the dynamics of averaged firing-rate models. This approach will be used in the result section to derive macroscopic limits and demonstrate the validity of the reduction. 
\subsection*{Neurons and networks}
We consider $P$ populations networks composed of $(N_{\alpha})_{\alpha\in\{1\cdots P\}}$ neurons. Functional structures involve very large numbers of neurons $N_{\alpha}$ in each populations, and similarly to the case of gas dynamics, we will be interested in the limit where all $N_{\alpha}$ tend to infinity (the \emph{mean-field limit}) in order to exhibit regularization and averaging effects. Each neuron $i$ in population $\alpha$ is described by the membrane potential $v^i_t$ (also called activity of a neuron in this paper) and additional variables gathered in a $d$-dimensional variable $Z^i_t$, representing for instance ionic concentrations in the Hodgkin-Huxley model, or a recovery variable in the Fitzhugh-Nagumo or McKean models. These variables satisfy a stochastic differential equation:
\begin{equation}\label{eq:NetworkGeneral}
	\begin{cases}
			dv^i_t&=\displaystyle{\bigg(F_{\alpha}(v^i,Z^i)\,+I^\alpha(t)+\sum_{j=1}^{N} J_{ij} v^j_t\ast h} \bigg)dt+\sigma_{\alpha}dW^i_t\\
			dZ^i_t&=G_{\alpha}(v^i_t,Z^i_t)\,dt+\Gamma_{\alpha}dB^i_t
	\end{cases}
\end{equation}
In this equation, the functions $F_\alpha$ and $G_\alpha$ describe the intrinsic dynamics of all the neurons in population $\alpha$. The parameters $\sigma_{\alpha}\in\R$ and $\Gamma_{\alpha}\in\R^{d\times d}$ describe the standard deviation of the noisy input received from each component of $X^i=(v^i,Z^i)$, classically driven by independent one (resp. $d$) dimensional Gaussian white noise $(W^i_t)$ (resp. $B^i_t$). $I^\alpha(t)$ represents the external input neurons of population $\alpha$ receive. The coefficients $J_{ij}$ are the synaptic weights of the connection $j\to i$. Spiking interactions between neurons are here modeled as the convolution of the presynaptic membrane potential with the impulse response of the synapse noted $h$, where $h(t) = \frac{1}{\tau_s} e^{-\frac{t}{\tau_s}} \mathds{1}_{t>0}$ with $\tau_s = 10\ ms$ is the characteristic time of the synapses and $\mathds{1}_{t>0}$ the Heaviside function. This is an idealization synapses dynamics under the assumption that the synapse is linear. 

Three main models used in computational neuroscience addressed in the present  manuscript are the McKean model~\cite{mckean:70}, the Fitzhugh-Nagumo model~\cite{fitzhugh:69} and the Hodgkin-Huxley model~\cite{hodgkin-huxley:52}. These models are parametrized so that the time unit is in milliseconds.

\paragraph{The McKean model}
is a piecewise continuous approximation of the Fitzhugh-Nagumo model allowing explicit calculations~\cite{mckean:70}. The membrane potential of neuron $i$ is written $v^i$ and its adaptation variable $w^i$. They satisfy the following network equations: 
\begin{equation}
  \left\{
    \begin{array}{l}
	 dv^i_t = \Big(f(v^i_t) - w^i_t + I^i + \sum_j J_{ij}\ \big(v^j_t \ast h\big)\Big)\,dt+\sigma^idW^i_t\\
	d{w^i_t} = \eps_w(v^i_t-w^i_t + b)dt
    \end{array}
\right.
\label{eq: MK network}
\end{equation}
with
\begin{equation*}
f(v) = 
\left\{\begin{array}{cc}
  -lv - (l + c)a & \mbox{if } v \leq -a\\
  c v & \mbox{if } a < v <a\\
  -lv + (l + c)a & \mbox{if } a \leq v \\
 \end{array}
\right.
\end{equation*}
and $\eps_w = 0.1$, $l = 1$, $a = 1$, $c = 0.5$, and  $b = 0.8$.

\paragraph{The Fitzhugh-Nagumo model}
was introduced in \cite{fitzhugh:69} and has been widely studied as a paradigmatic excitable systems for its simplicity and ability to produce spiking behaviors. This model was introduced in order to allow analytical study in a simpler model  mimicking the essential features of the Hodgkin-Huxley neuron. 
Fitzhugh-Nagumo models describe the activity of the membrane potential $v^i$ and recovery variable $w^i$ of neuron $i$ in the network through the equations:
\begin{equation}\label{eq:FhNNetwork}
  \left\{
    \begin{array}{l}
	 dv^i_t = \Big(v^i_t - \frac{(v^i_t)^3}{3} - w^i_t + I^i + \sum_j J_{ij}\ \big(v^j_t \ast h\big)\Big)\,dt+\sigma^idW^i_t\\
	d{w^i_t} = \eps_w(v^i_t- a w^i_t + b)dt
    \end{array}
\right.
\end{equation}
with $\eps_w = 0.08$, $a = 0.8$ and $b = 0.7$.

\paragraph{The Hodgkin-Huxley model: }
Probably the most relevant neuron model from the electrophysiological viewpoint, the Hodgkin-Huxley model describes the evolution of the membrane potential in relationship with the dynamics of ionic currents flowing across the cellular membrane of the neuron, developed in~\cite{hodgkin-huxley:52} after thorough observation of the giant squid axon revealed the prominent role of potassium and sodium channels for excitability, and leak chloride currents. Networks of Hodgkin-Huxley neurons are described by the stochastic differential equation:
\begin{equation}\label{eq:HHNetwork}
  \left\{
    \begin{array}{ll}
	C dv^i_t &= \Big(I^i - g_K (n^i)^4(v^i_t - E_K) - g_{Na}(m^i)^3 h^i(v^i_t - E_{Na}) \\
	& \qquad - g_L(v^i_t - E_L) + \sum_j J_{ij}\ \big(v^j_t \ast h\big)\Big)dt + \sigma^i dW_t^i\\
	dn^i_t &= \big(\alpha_n(v^i_t)(1 - n^i) - \beta_n(v^i_t)n^i\big) dt\\
	dm^i_t &= \big(\alpha_m(v^i_t)(1 - m^i) - \beta_m(v^i_t)m^i\big) dt\\
	dh^i_t &= \big(\alpha_h(v^i_t)(1 - h^i) - \beta_h(v^i_t)h^i\big) dt\\
    \end{array}
\right.
\end{equation}
where
\begin{align*}
 \alpha_n(v) &= 0.01 \frac{10-v}{\exp(\frac{10-v}{10}) - 1}\quad , \quad  &\alpha_m(v) = 0.1 \frac{25-v}{\exp(\frac{25-v}{10}) - 1}\quad , \quad  \alpha_h(v) &= 0.07 \exp(\frac{-v}{20})\\
 \beta_n(v) &= 0.125 \exp(\frac{-v}{80}), &\beta_m(v) = 4. \exp(\frac{-v}{18}) \quad , \quad  \beta_n(v) &= \frac{1}{\exp(\frac{30-v}{10}) + 1}
\end{align*}
and\footnote{Parameters are adjusted so that the resting state of the membrane potential $v$ is equal to $0 mV$.} $C= 1\mu F/cm^2$, $g_K = 36mS/cm^2$, $g_{Na} = 120 mS/cm^2$, $g_L = 0.3 mS / cm^2$, $E_K = -12mV$, $E_{Na} = 120mV$ and $E_L = 10.6 mV$. The dynamics of this system shows deep non-linear intricacies even in the case of deterministic, single-neuron system.

\subsection*{Mean-field limits}
These network equations are extremely complex to analyze and simulate. However, in the mean-field limit, one can access to the asymptotic behavior of neurons in large networks. We consider that synaptic weights are heterogeneous and have a typical mean and standard deviation depending on the populations they belong to, e.g. $J_{ij}\sim \mathcal{N}(\frac{J_{\alpha\gamma}}{N_{\gamma}},\frac{\lambda}{N_\gamma})$ where $J_{\alpha\gamma}$ is the typical connectivity weight between populations $\alpha$ and $\gamma$ and $\lambda$ the synaptic weights disorder. In this case, the network is described by a set of delayed stochastic differential equations, and the propagation of chaos applies, as shown in~\cite{touboul2011mean}. This means that all the neurons in the same population (which we call $\alpha$) follow the same law. This law is described by the following mean-field equation
\begin{equation}\label{eq:MFEGeneral}
	\begin{cases}
			dv^{\alpha}_t&=\bigg(F_{\alpha}(v^{\alpha},Z^{\alpha})\,+\sum_{\gamma=1}^P J_{\alpha\gamma} \Exp{v^{\gamma}_t}\ast h +I^\alpha(t)\bigg)dt+\sigma_{\alpha}dW^{\alpha}_t\\
			dZ^{\alpha}_t&=G_{\alpha}(v^{\alpha}_t,Z^{\alpha}_t)\,dt+\Gamma_{\alpha}dB^{\alpha}_t
	\end{cases}
\end{equation}
Observe that this equation, describing the process governing each neuron in population $\alpha$, is very similar to the network equation \eqref{eq:NetworkGeneral}. Only the interaction term is different since it involves the computation of an expectation of the activity. The mathematical study of this type of equation is generally extremely complex: in particular, it is not a stochastic differential equation, but rather an implicit equation in the set of stochastic processes. In our present approach, we will manage to bypass this difficulty by considering the macroscopic activity of a population.

\subsection*{Firing Rates, Macroscopic Activity and Dynamics}
The probabilistic behavior of neurons in the limit $N\to \infty$ being characterized, we can define macroscopic variables caracterizating the state of the population.

The firing-rate is usually considered a good candidate as a macroscopic descriptor of the population activity. Heuristically, its corresponds to the average number of spikes fired in a certain time window. However, counting spikes is a non-linear and non trivial operation which is difficult to handle mathematically.

We will rather use a simpler definition of the macroscopic activity: the average within a population and during a certain time of the membrane potential of neurons. This definition has the interesting property of being a linear transformation of the activity of neurons. It is also closely related to the firing-rate. Indeed, given that neurons communicate via spikes which are stereotyped electrical impulses of extreme amplitude, integrating the value of the membrane potential during a time window and dividing by the area under a spike provide a rough estimate of the number of spikes emitted. However, the two measures are not exactly similar: this definition of macroscopic activity is impacted by the activity under the firing threshold of neurons as opposed to pure firing-rates.

More precisely, we define the \textit{macroscopic activity} $\nu^{\alpha}$ of population $\alpha$ as the spatial mean of the membrane potential $v^{\alpha}$ over the neurons in the populations and temporally convolved with the ``time window'' function $g$ of width $\theta = 100\ ms$\footnote{Concretely, $g(t) = \frac{1}{K} e^{-\frac{t^2}{s^2}}$ with $s$ chosen so that $K g(\theta /2) = 0.01$ and $K$ so that $\int_\R g(t)dt = 1$.} larger than the duration of the spikes, but small enough to resolve fine temporal structure of the network activity. The propagation of chaos property ensures that all neurons are independent; this allows to identify the population-averaged voltage with the statistical expectation of the voltage variable in the mean-field limit and leads to the following definition of the macroscopic activity:
\begin{equation}\label{eq:FiringRateGeneral}
\nu^{\alpha}(t) \eqdef \big(\E(v^{\alpha}) \ast g\big)(t) 
\end{equation}
Thanks to the linearity of synapses, using this definition together with the mean field equation \eqref{eq:MFEGeneral} leads to the following caracterization of the macroscopic activity.
\begin{equation}
 \dot{\nu}^{\alpha}=\Exp{F_{\alpha}(X^{\alpha})}\ast g + \sum_{\beta=1}^P J_{\alpha \beta} \big(\nu^{\beta} \ast h\big) + \tilde{I}^\alpha(t) 
 \label{eq: MFE non-closed}
\end{equation}
where $X^{\alpha}=(v^{\alpha},Z^{\alpha})$, $\tilde{I}^{\alpha}$ denotes $I^{\alpha}\ast g$.

This equation is not closed because of the term $\Exp{F_{\alpha}(X^{\alpha}_t)}\ast g$ which is not expressed as a function of the macroscopic activity. Because of the nonlinearity of $F_{\alpha}$, it is not likely that this quantity only depends on ${\nu}^{\alpha}$. Moreover, the membrane potential depends on additional variables and the macroscopic activity hence involve expected value of functions of these variables convolved with the time kernel $g$.

\subsection*{Computing the effective non-linearity}
We now attempt to make sense of the term $\Exp{F_{\alpha}(X^{\alpha}_t)}\ast g$ to find a closed formula on the macroscopic activity. We introduce the ansatz that this term can be written as the sum of a linear functional of the macroscopic activity and a non-linear term applied to the effective input to a population (including the synaptic connections). Defining this effective input $x^{\alpha}$ as:
\begin{equation}
x^{\alpha} \eqdef \sum_{\beta=1}^P J_{\alpha \beta} \big(\nu^{\beta} \ast h\big) + \tilde{I}^\alpha(t) 
\label{eq: effective input}
\end{equation}
the ansatz reads:
\begin{equation}
 \Exp{F_{\alpha}(X^{\alpha}_t)}\ast g = L(\nu^{\alpha}) + S(x^{\alpha})
 \label{eq: Sigmoid def 1}
\end{equation}
where $L$ is a linear functional and $S$ is a non-linear mapping which remains to be determined. The choice of this ansatz was motivated by the analytical treatment of networks of deterministic McKean neurons (see appendix), which naturally exhibits a relation like \eqref{eq: Sigmoid def 1}. The choice of the linear functional $L$ is dictated by the neuron model used. In general, the function $S$ has to be computed numerically.

To evaluate $S$, one need to assume that both the inputs and the synapses are slow compared to the dynamics of the neurons. Thus, the effective input $x^\alpha$ can be considered as constant during the time the neurons reach an asymptotic regime related to that input state. If this regime is stationary, the value function $S$ at $x^\alpha$ will be the average of $F_{\alpha}- L$ applied to that stationary stochastic process, and therefore will provide a quantity only depending on $x^{\alpha}$. If the regime is periodic in law, then taking a time window $g$ of size $\theta$ larger than the period\footnote{we recall that $\theta$ was assumed to be small enough to resolve the temporal structure of the neuron's activity.} will also yield a constant value for our nonlinear term. Because the mean-field equation \eqref{eq:MFEGeneral} and a single neuron equation only differ in the interaction term which is assumed constant here, the computation of $S(x^\alpha)$ simply corresponds to computing the temporal average of 
$F_{\alpha}- L$ along the trajectory a single noisy neuron forced with a constant input $x^\alpha$. Actually, for readability reasons\footnote{This is motivated by the 
combination of equations \eqref{eq: MFE non-closed}, \eqref{eq: effective input} and \eqref{eq: Sigmoid def 1} into equation \eqref{eq: averaged network theoretic}} we will rather focus on the function $\tilde{S}$ which is simply defined as $\tilde{S}(x^\alpha) = S(x^\alpha) + x^\alpha$. 

\begin{figure}[ht]
	\centering
		\subfigure[Regime I neuron]{\includegraphics[width=.3\textwidth]{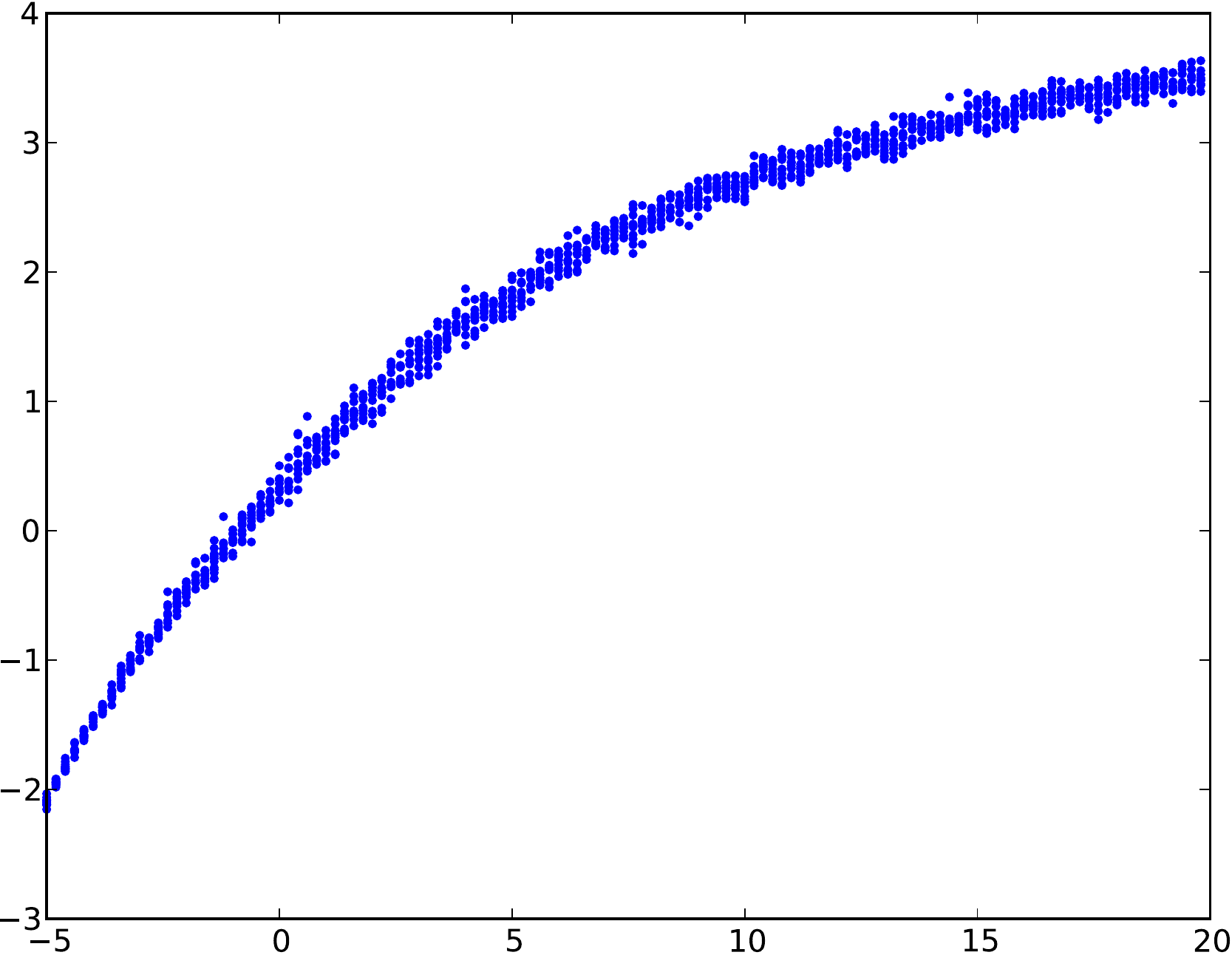}}\qquad\qquad
		\subfigure[Regime II neuron]{\includegraphics[width=.3\textwidth]{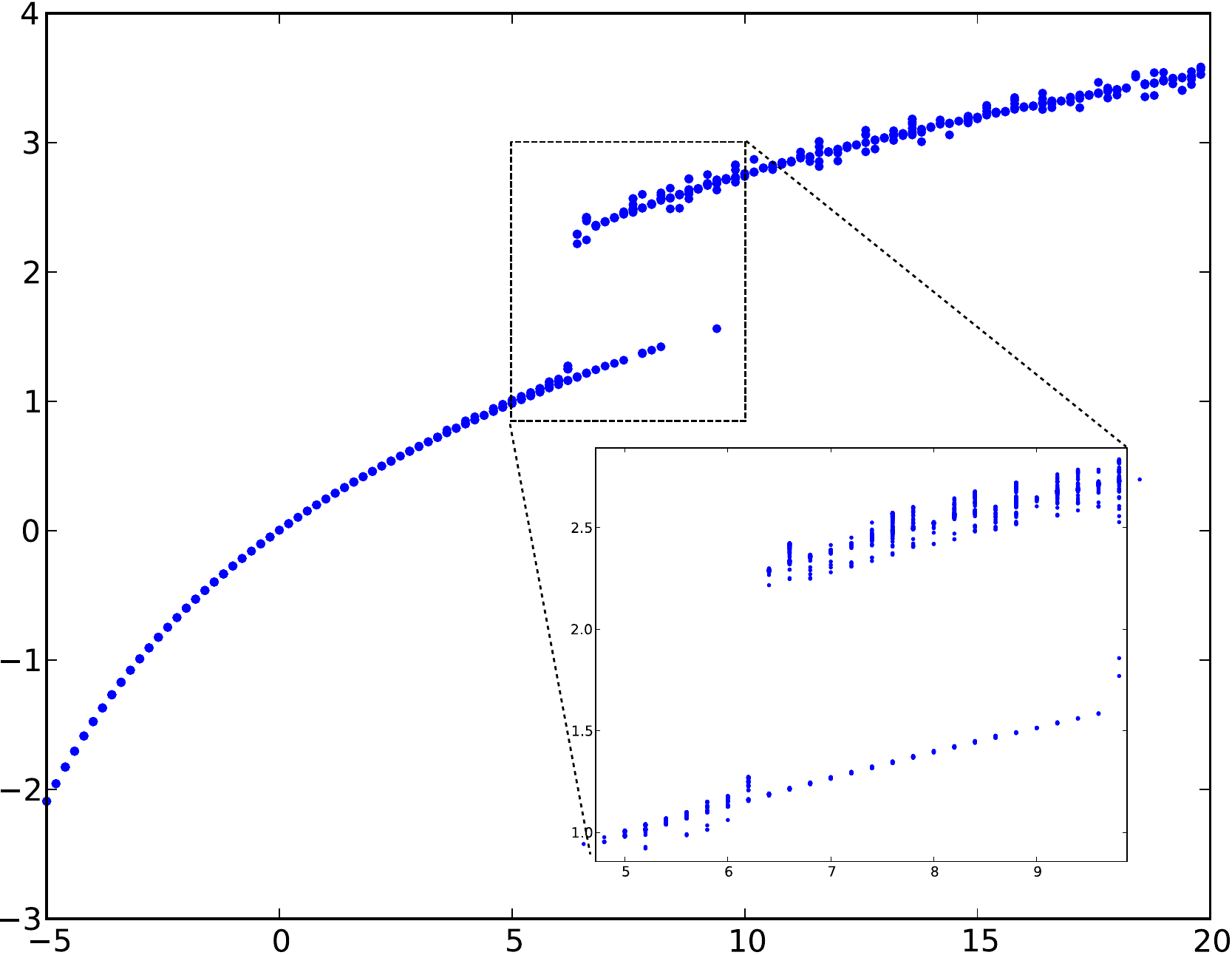}}
	\caption{Value of $\tilde{S}(x^{\alpha})$ computed for 10 different initial conditions (and 200 in the inset of (b)) for the Hodgkin-Huxley model with noise $\sigma=2$ (a) and $\sigma=0$ (b). This shows that the level of intrinsic noise change the regime of a neuron.}
	\label{fig:PointsSigmoid}
\end{figure}

One of the pitfalls of this methods occurs if the behavior of the mean-field equation depends on the initial condition and present multiple stable stationary or periodic attractors, the quantity $\Exp{F_{\alpha}(X^{\alpha}_t)}\ast g$ can take different values depending on the initial condition. A neuron model (together with a particular set of parameters) will be said to be of regime I if there is only one of these attractor for any initial condition. Similarly, if there are p-attractors, the neuron model is said to be of regime p. Figure \ref{fig:PointsSigmoid} shows different values for $\tilde{S}(x^\alpha)$ when starting from different initial condition. Figure \ref{fig:PointsSigmoid}.a) shows the solution for regime I and figure \ref{fig:PointsSigmoid}.b) for regime II. The points obtained (see Fig.~\ref{fig:PointsSigmoid}) are then segmented into a few clusters (in our case, one or two in regimes I and II respectively) and smoothed out into a surface (or a union of surfaces) see figure \ref{fig:SigmoidsNoise}. This procedure is relatively time consuming. The result of extensive simulations on the McKean, Fitzhugh-Nagumo and Hodgkin-Huxley models, using this numerical procedure, are freely available online as well as the algorithm generating these data. 

\subsection*{Linear part $L$ for the different models}
Identifying $L$ for a given model, consists in gathering the most linear terms in the intrinsic dynamics of a single neuron $F$. This linear terms can be time-delayed. A careful study of the different models suggests that the linear parts for the different neurons can be chosen as in table \ref{table: linear part}.
\begin{table}[htbp]
\begin{center}
    \begin{tabular}{| c | l | }
    \hline
    Model & Linear part \\ \hline
    McKean & $L(\nu^\alpha) = -\nu^{\alpha} \ast (l \delta + h_{\tau_w})$ \\ \hline
    Fitzhugh-Nagumo & $L(\nu^\alpha) = -\nu^{\alpha} \ast (\frac{4}{3} \delta + h_{\tau_w})$\\ \hline
    Hodgkin-Huxley & $L(\nu^\alpha) = - g_L \nu^\alpha$\\
    \hline
    \end{tabular}
\end{center}
\caption{Linear part $L$ for the different models. $\delta$ is the Dirac function centered at $0$ and $h_{\tau_w} = \eps_w e^{-\eps_w t}\mathbbm{1}_{t>0}$ for the first two models.}
\label{table: linear part}
\end{table}

For McKean and Fitzhugh-Nagumo models, the convolution accounts entirely for the existence of the adaptation variable $w$ which will therefore still be present in the reduced model. The choice of the linear part of the McKean model can be understood better by reading the appendix where it naturally comes out of the computations. The choice of the coefficient $4/3$ for the Fitzhugh-Nagumo model relies on an analogy to the McKean neuron. It corresponds to the (absolute value) of the slope of the straight line approximating the negative decreasing part of the non-linearity Fitzhugh-Nagumo $v - \frac{v^3}{3}$. As for the Hodgkin-Huxley, we were not able to extract a linear delayed term in the intrinsic dynamics of a neuron so we simply chose the linear decay already present in the equations.

\subsection*{Simulation of the macroscopic equations}
For regime I neurons, when the effective non-linearity $\tilde{S}$ is univalued, simulations of the macroscopic equations simply reduce to solving numerically the following ordinary differential equation
\begin{equation}
\dot{\nu}^{\alpha}=L(\nu^{\alpha}) + \tilde{S}\Big(\sum_{\beta=1}^P J_{\alpha \beta} \big(\nu^{\beta} \ast h\big) + \tilde{I}^\alpha(t) \Big) 
\label{eq: averaged network theoretic}
\end{equation}

\noindent For regime II neurons, when the initial condition is not in the bistable region, we will consider that the averaged system pursues on the initial attractor (fixed point or spiking cycle) when possible, and switches attractors if the activity brings the system in regions where the initial attractor disappears. In details, let us denote by $\tilde{S}(x^\alpha,1)$ the branch of stable fixed points, defined as long as $x^\alpha\leq I_H$, and by $\tilde{S}(x^\alpha,2)$ the branch of periodic orbits defined for $x^\alpha \geq I_{FLC}$. The macroscopic activity of population $\alpha$ in a $P$-population network hence satisfies the equations:
\begin{equation}\label{eq:HHRed}
	\begin{cases}
		\dot{\nu^{\alpha}}=L(\nu^\alpha) + \tilde{S}\Big(\sum_{\gamma=1}^P J_{\alpha\beta}(\nu^{\beta}*h) + \tilde{I}^{\alpha}, p(t) \Big)\\
		\dot{p(t)}= \delta_{x^\alpha-I_H}\delta_{p,1}-\delta_{x^\alpha-I_{FLC}}\delta_{p,2}.
	\end{cases}
\end{equation}

\section*{Results}
In this section, we evaluate the accuracy of the reduced model presented above for the three neuron models McKean, Fitzhugh-Nagumo and Hodgkin-Huxley. First, we address the computation of the effective non-linearity both in the deterministic and noisy cases. Second, we confront the time course of the macroscopic activity calculated according to our reduction against the a posteriori average of the activity of a spiking network.

\subsection*{Effective non-linearity without noise}
\begin{figure}[htbp]
	\centering
		\subfigure[McKean bifurcation diagram]{\includegraphics[width=.3\textwidth]{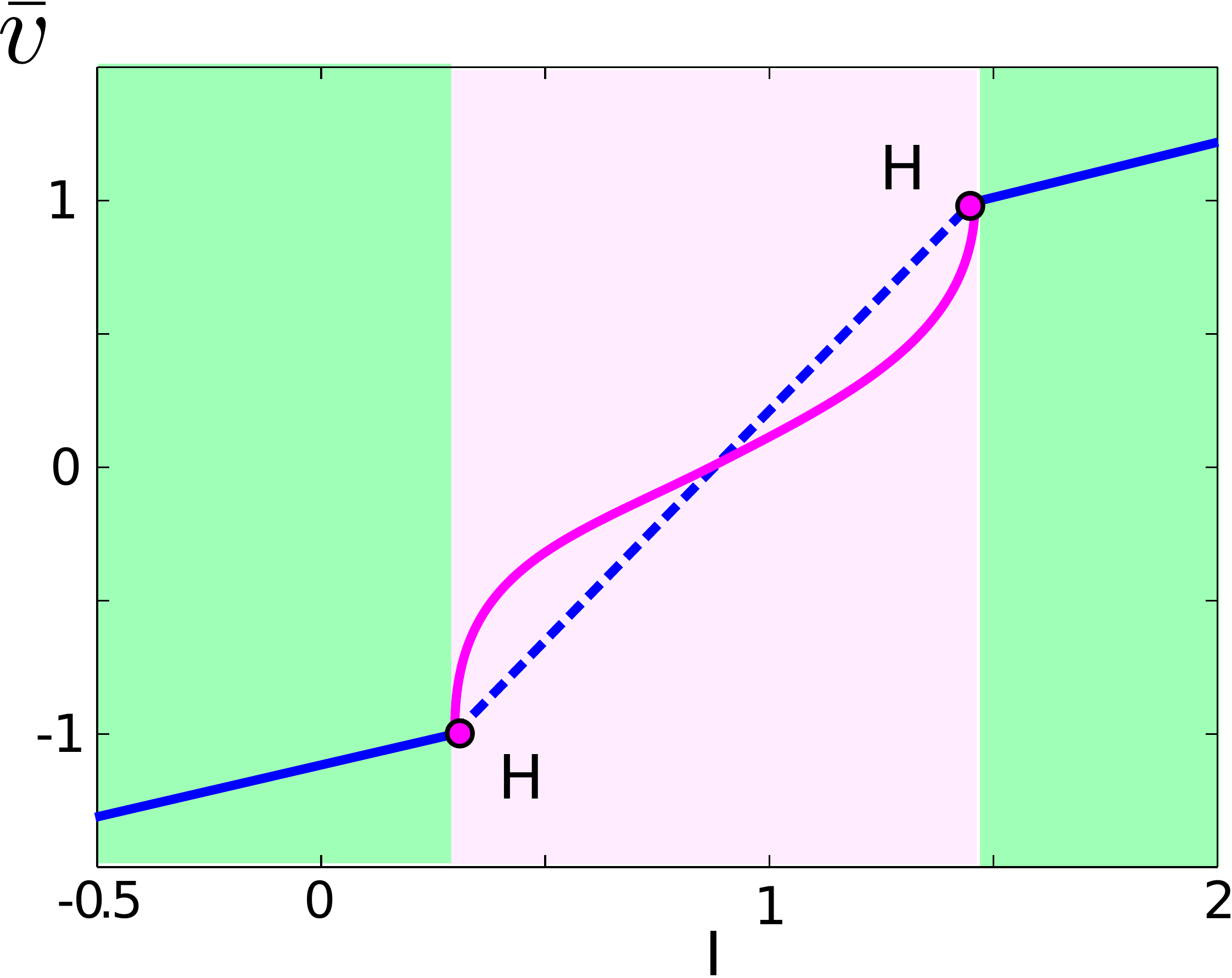}\label{fig:MKBifs}}\quad
		\subfigure[FitzHugh-Nagumo bifurcation diagram]{\includegraphics[width=.3\textwidth]{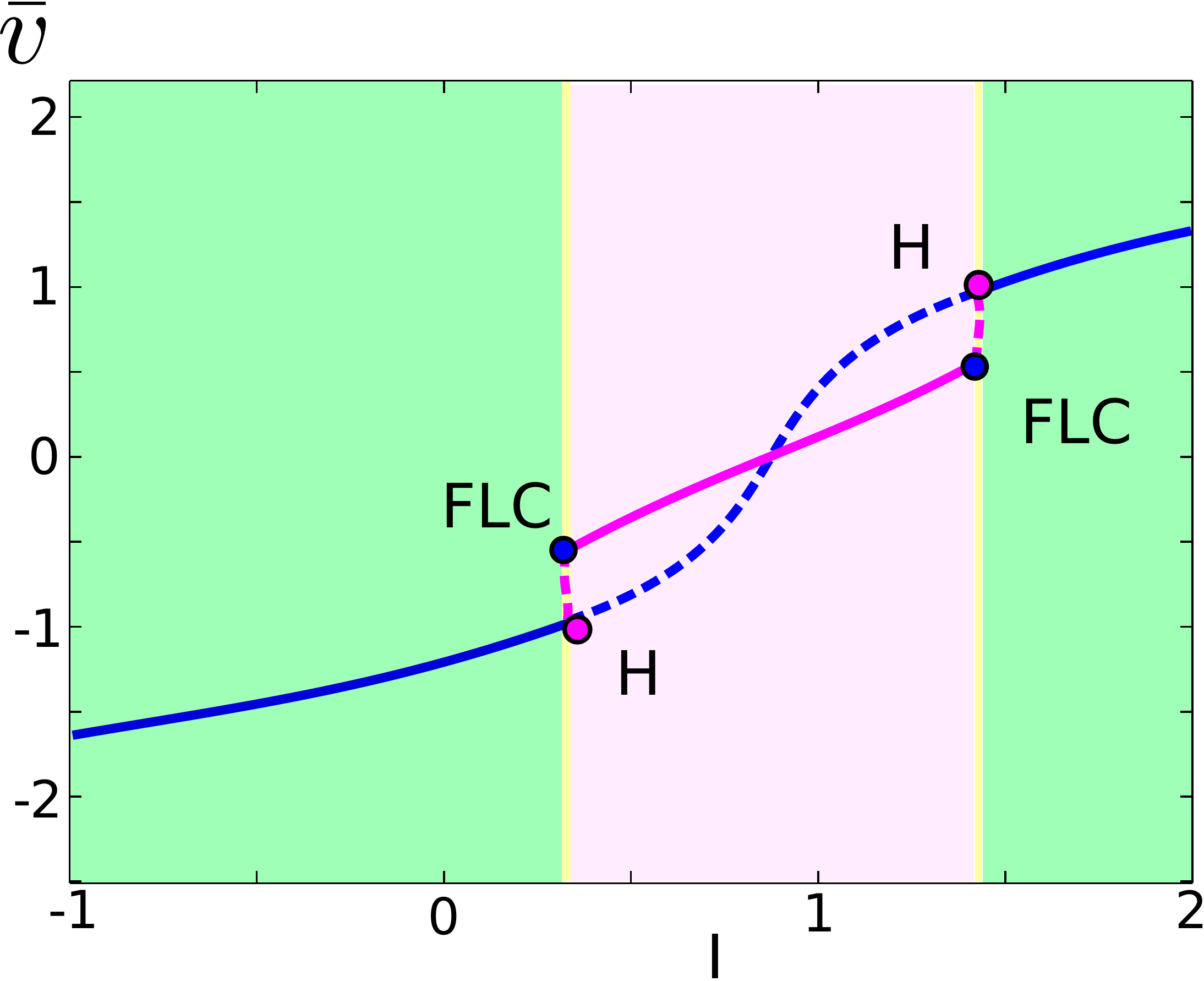}\label{fig:FhNBifs}}\quad
		\subfigure[Hodgkin-Huxley bifurcation diagram]{\includegraphics[width=.3\textwidth]{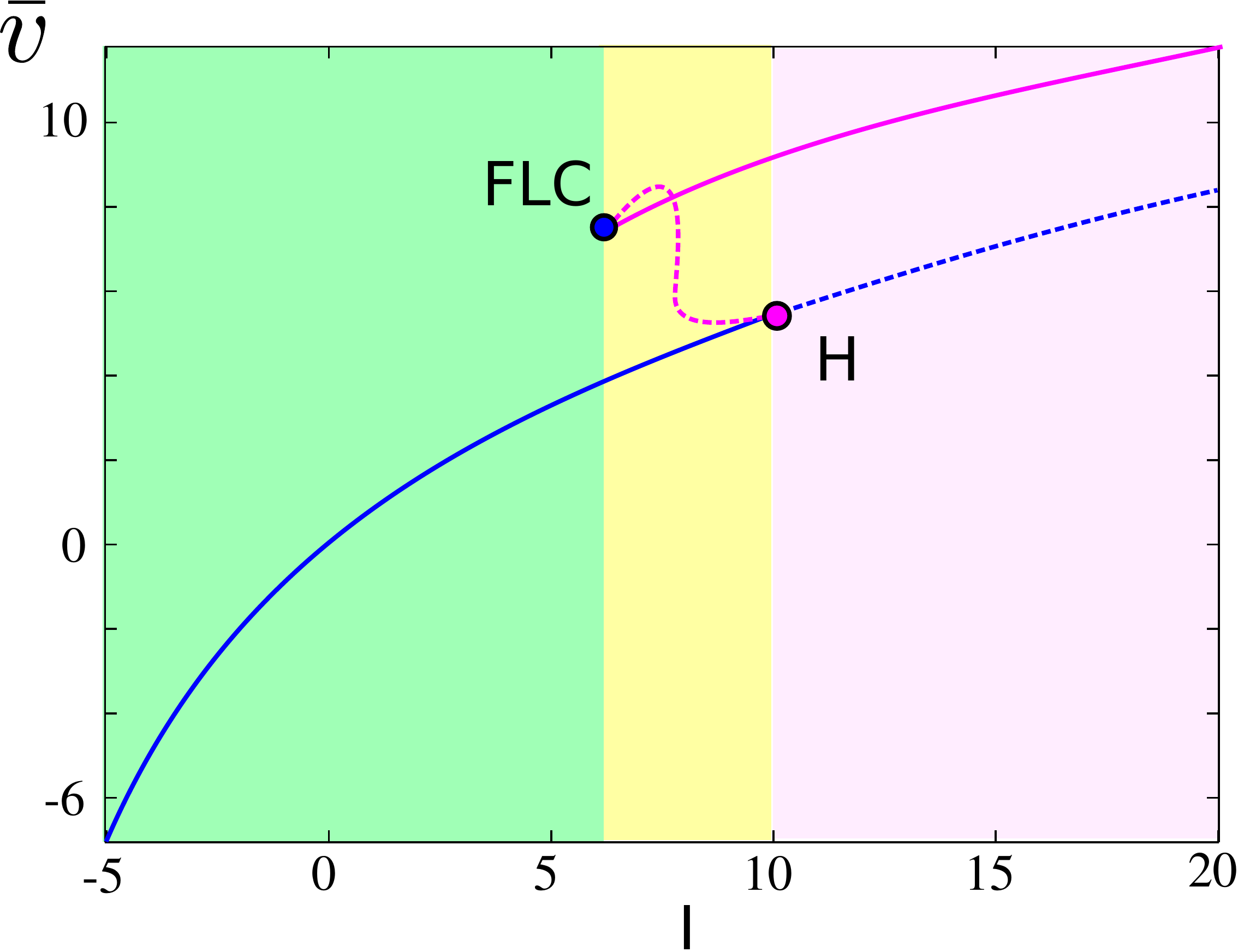}\label{fig:HHBifs}}\\
		\subfigure[McKean frequencies]{\includegraphics[width=.3\textwidth]{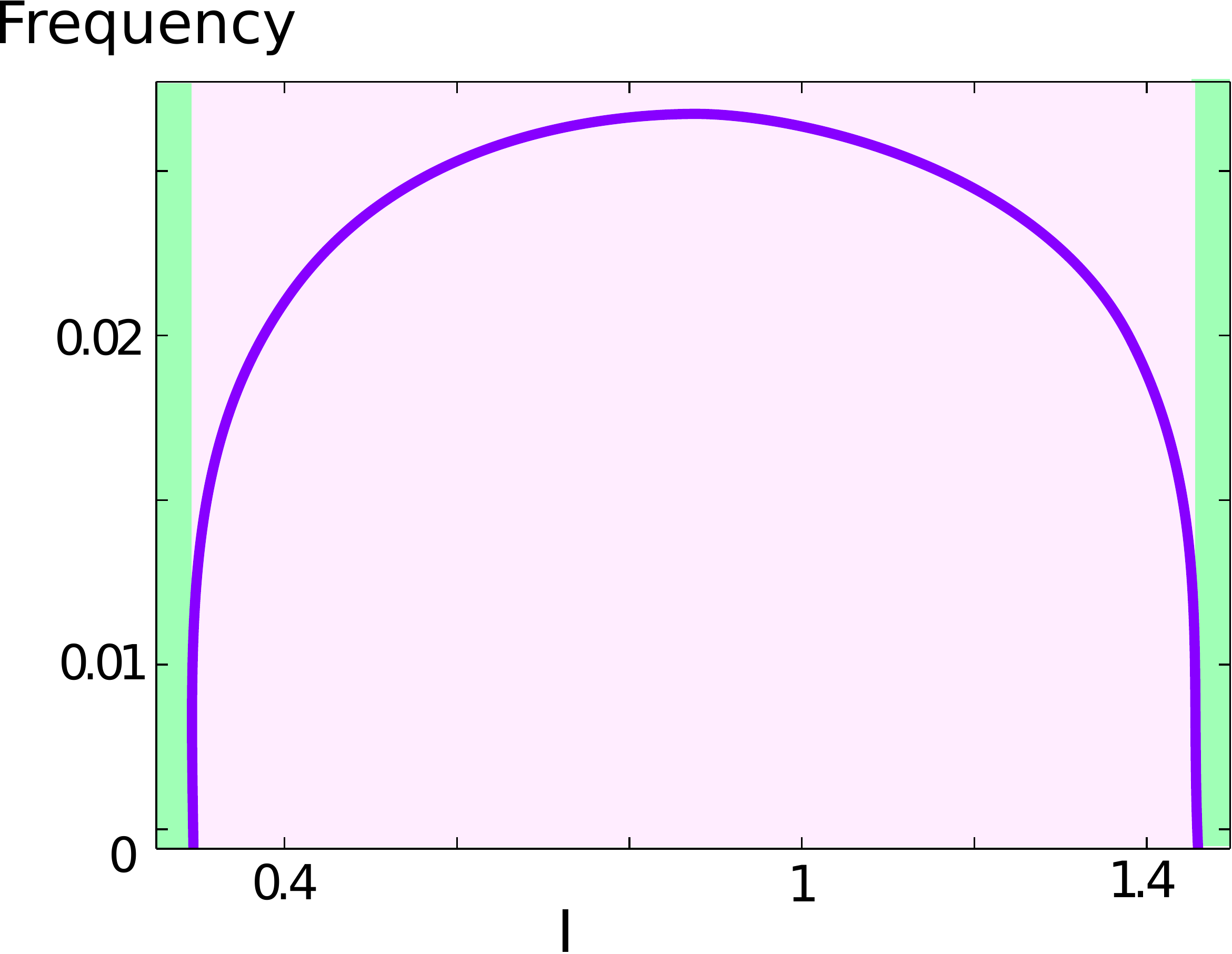}}\quad
		\subfigure[FitzHugh-Nagumo frequencies]{\includegraphics[width=.3\textwidth]{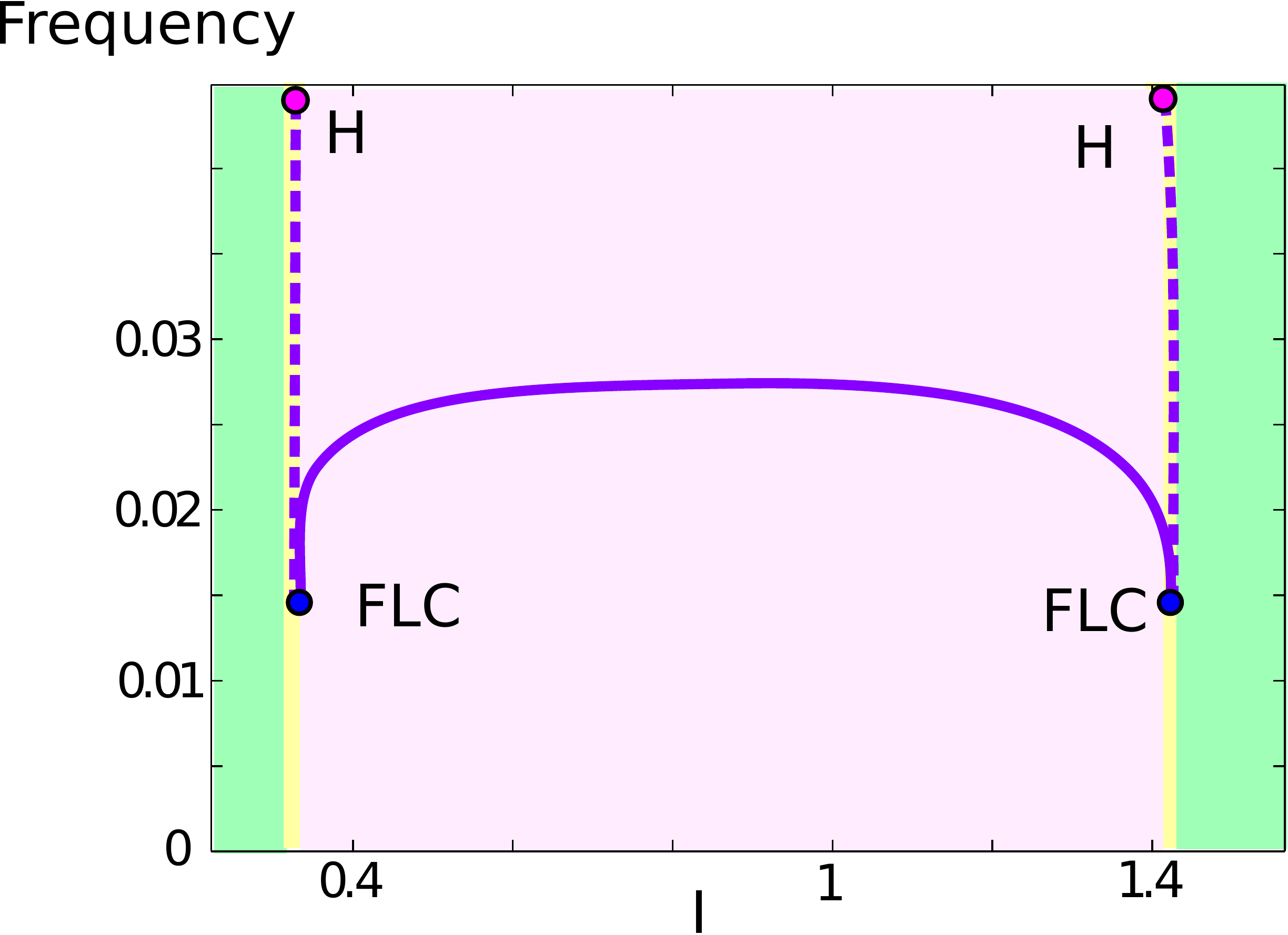}}\quad
		\subfigure[Hodgkin-Huxley frequencies]{\includegraphics[width=.3\textwidth]{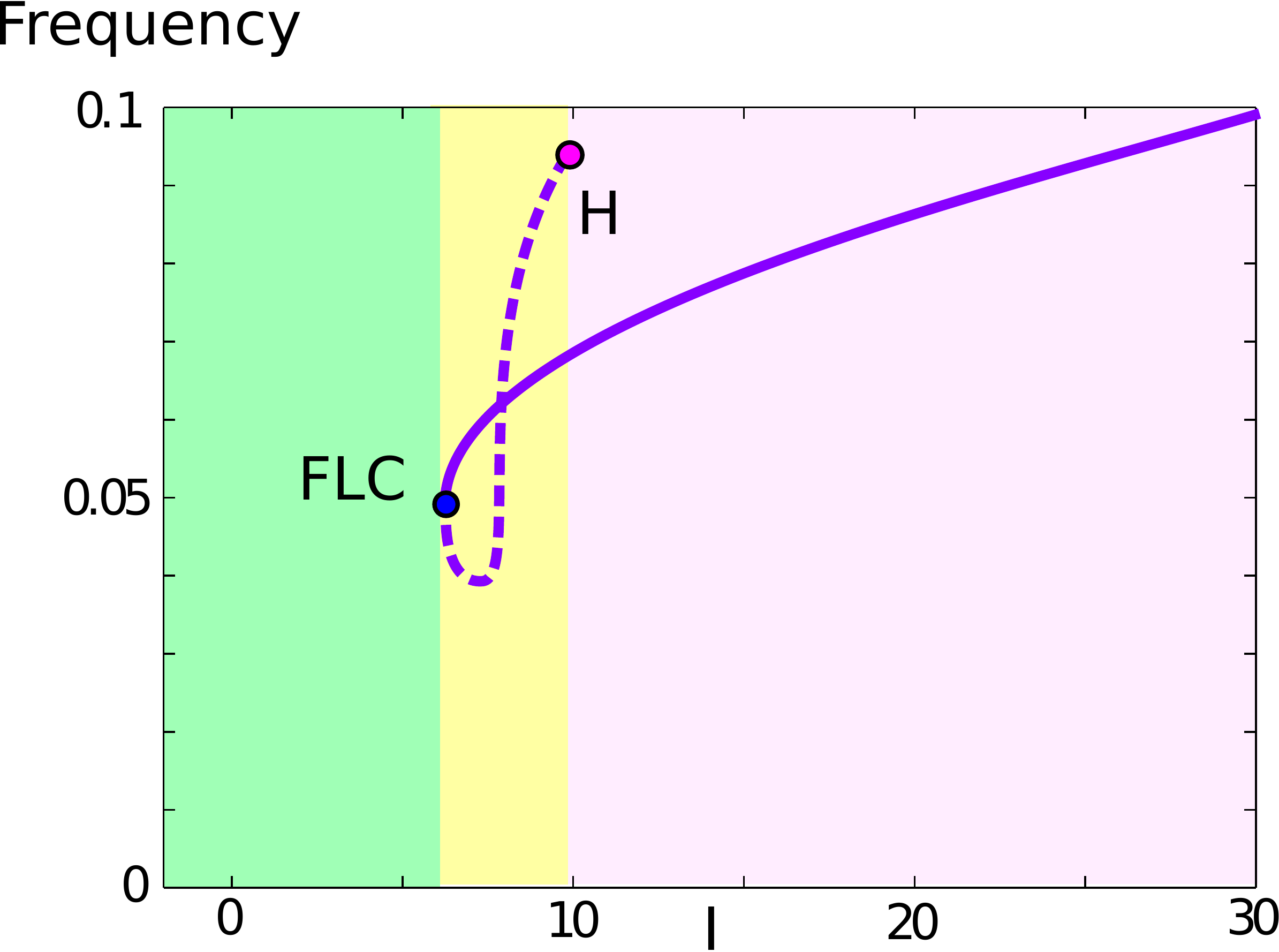}\label{fig:HHfreqs}}
	\caption{Bifurcation diagrams (upper row) representing the temporal average of the solutions (i.e. the fixed points and average value in the case of periodic orbits) and frequency of the regular spiking regime (lower row) in the McKean model (left), Fitzhugh-Nagumo model (center) and Hodgkin-Huxley model (right). }
	\label{fig:BifDiag}
\end{figure}

In the deterministic reduction (small noise limit), the effective non-linearity can be understood through the bifurcation analysis of a single neuron. Indeed, the effective sigmoid amounts to computing the temporal average $\bar{v}$ of the voltage solution of the deterministic single-cell system upon variation of the input. The result of that analysis using the numerical software XPPAut~\cite{ermentrout1987simulating} is displayed in Figure Fig.~\ref{fig:BifDiag}. In these diagrams, we colored regions of stationary solutions (green), periodic solutions (purple) and bistable regimes with co-existence of a stable stationary solution and a stable periodic orbit. 

\subsubsection*{The McKean neuron}
Although the deterministic McKean neuron is analytically treated in the appendix, we now consider it under the angle of bifurcations for consistency with the other models. In the McKean neuron, the non-differentiable, piecewise-continuous nature of the flow gives rise to a non-smooth Hopf bifurcation associated with a branch of stable limit cycles. The emergence of the cycle arises through a non-smooth homoclinic bifurcation, hence corresponding to the existence of arbitrarily slow periodic orbits, typical of a class I excitability in the Hodgkin classification\footnote{Observe that Hodgkin classification is different from the classification in regime we propose in this paper}. In this model, an important distinction is the absence of bistable regime: the average variable has a unique value whatever the initial condition (except singularly chosen equal to the unstable fixed point) and whatever the input chosen. In the present case, stable permanent regimes are unique. Therefore, there exists a single-valued 
function $\tilde{S}$ defining the dynamics of the macroscopic activity through equation \eqref{eq: averaged network 
theoretic}.

\subsubsection*{The Fitzhugh-Nagumo model}
The bifurcation diagram of the Fitzhugh-Nagumo neuron as a function of the input level (see figure~\ref{fig:FhNBifs}) presents multi-stability. For small negative input, the system presents a stable fixed point. Increasing the value of the input makes the fixed point lose stability through subcritical Hopf bifurcation, and unstable limit cycles appear. These limit cycles undergo a fold and a branch of limit cycles appear, overlapping in a small parameter region the state where a stable equilibrium exists. This small parameter region again corresponds to a bistable regime with co-existence of a resting state and of a regular spiking behavior. This branch of stable limit cycles corresponds to a regular spiking regime. As the input is further increased (in a biologically unplausible range), the same scenario arises symmetrically: the branch of stable periodic orbits undergoes a fold of limit cycles, a branch of unstable periodic orbits emerges from this bifurcation and connects with the unstable fixed point at 
a subcritical Hopf bifurcation and the unstable fixed point gains stability. Here again, the neuron corresponds to a class II excitability and a regime II, but the system could be well approximated by a excitability I/ regime I since the bistability region is of very small extent and the periodic orbits have very small periods when appearing, close from a class I excitability. Thus, we consider that the macroscopic activity evolves according to equation \eqref{eq: averaged network theoretic} as illustrated in the following.

\subsubsection*{The Hodgkin-Huxley model}
This model displays qualitatively the same behavior as the Fitzhugh-Nagumo model but with significant quantitative differences: the bifurcation diagram of the Hodgkin-Huxley neuron (see figure ~\ref{fig:HHBifs}) also displays multi-stability. We observe the existence of a branch of stable fixed points (green region) that undergo subcritical Hopf bifurcation for $x^\alpha = I_H$, associated to a family of unstable limit cycles (pink dotted lines represent the average value of $v$ along the cycle). This family of unstable limit cycles connects with a branch of stable limit cycles (pink solid line) through a fold of limit cycles bifurcation. These stable limit cycles are the unique attractor for large input (implausibly large input values will nevertheless see these cycles disappear in favor of a high voltage fixed point). The system presents a bistable regime (yellow input region) where a stable fixed point and a stable periodic orbit co-exist. The frequency along the cycle (Fig.~\ref{fig:HHfreqs}) shows a 
class II excitability in the Hodgkin classification: oscillations appear with a finite period and a non-zero frequency. The hysteresis present in the yellow region corresponds to what we called a regime II. In simulations, when the initial condition is not in the bistable region, we will consider that the system pursues on the initial attractor (fixed point or spiking cycle) when possible, and switches attractors if the activity brings the system in regions where the initial attractor disappears, as explained in the Material and Methods section. This method is chosen here because the bistability only appears for small noise, regimes in which switches between the different attractors are rare.

\subsection*{Effective non-linearity with noise}
\begin{figure}[htbp]
	\centering
		\includegraphics[width=\textwidth]{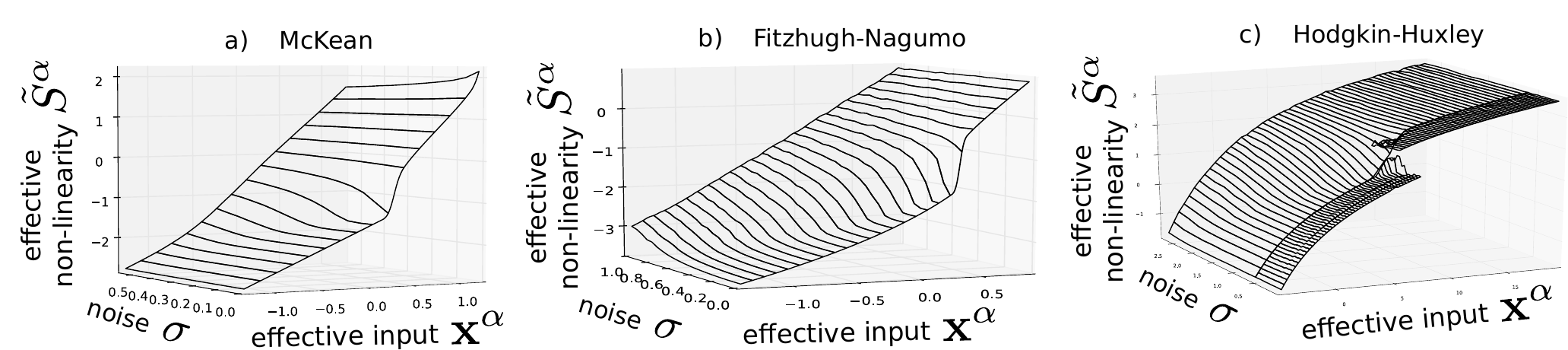}
	\caption{Effective non-linearities in the McKean, Fitzhugh-Nagumo and Hodgkin-Huxley model. Observe that noise tends to have a smoothing effect on the sigmoids.For the Hodgkin-Huxley model, we have empirically chosen a noise threshold under which the neuron was considered regime II and above which it is regime I. There are thus 2 branches below the threshold and only one above.}
	\label{fig:SigmoidsNoise}
\end{figure}

When considering noisy input, no analytical approach could be performed. We use the method described in Material and Methods and plot the results in Figure~\ref{fig:SigmoidsNoise} where the effective non-linearity is displayed as a surface obtained as a function of noise and effective input. It appears relatively clear in the figure that noise tends to have a smoothing effect on the sigmoids. This effect is particularly clear in the Hodgkin-Huxley model where a multivalued function (regime II) is turned into a single valued smooth function (regime I).
In the case of the Fitzhugh-Nagumo model, we observe that the regime II is not observed in simulations in the presence of noise. This is due to the smallness of the parameter region corresponding to the bistable regime, and the averaged system can be well approximated by regime I dynamics. In the case of the Hodgkin-Huxley network, there are clearly two different behaviors depending of the level of intrinsic noise. When the noise is small (resp. large) the neuron is regime II (resp. I). Interestingly, this shows how a strong noise can qualitatively simplify the macroscopic dynamics of a network.

\subsection*{Comparison between reduced model and averaged spiking network}
We now simulate large networks of McKean, Fitzhugh-Nagumo and Hodgkin-Huxley neurons, compute numerically their macroscopic averaged activity, and compare the dynamics of this variable to simulations of the reduced ordinary differential equation involving our effective non-linearity function in all three cases.

\begin{figure}[htbp]
	\centering
		\subfigure[Inputs]{\includegraphics[width=.3\textwidth]{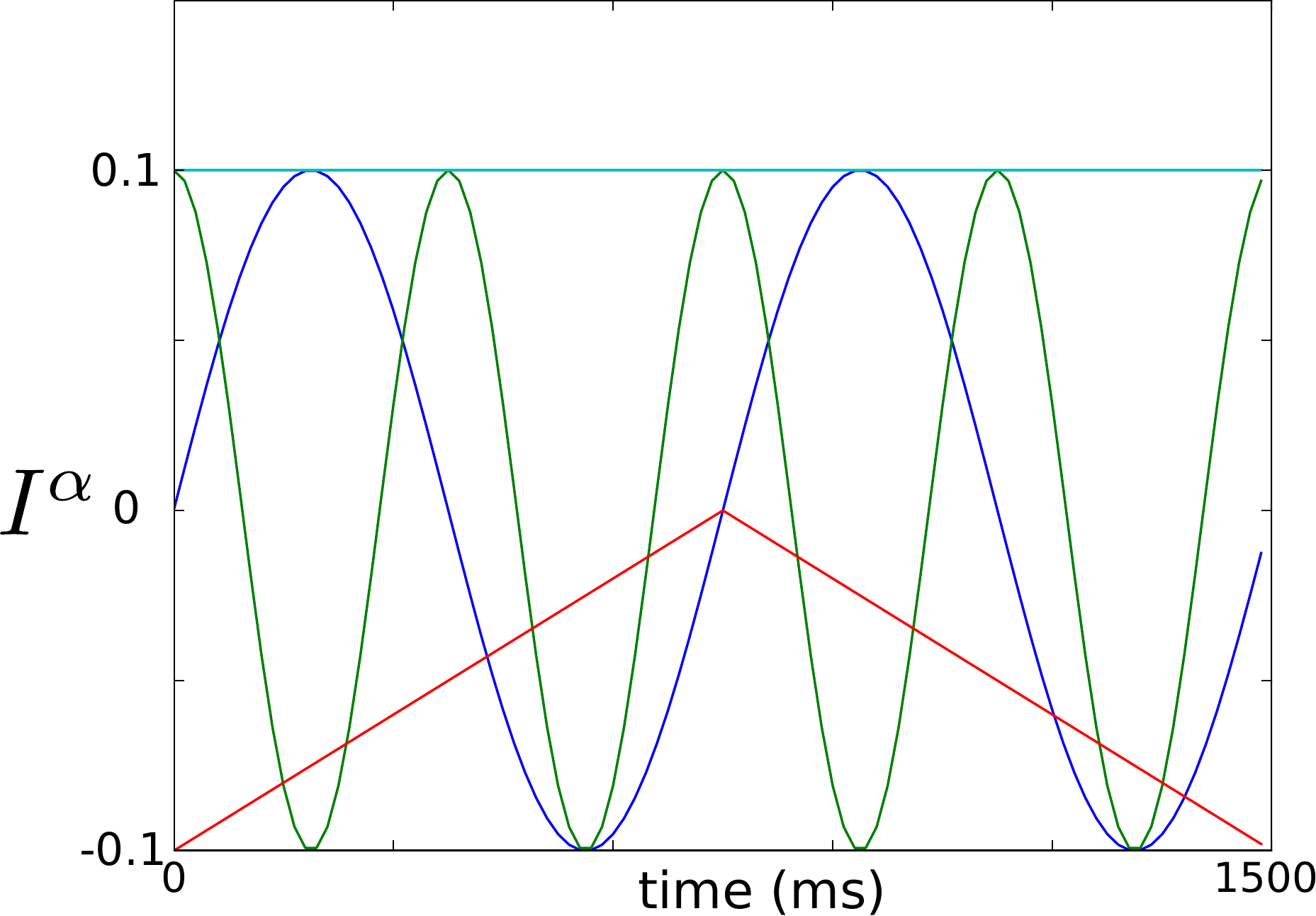}\label{fig: inputs}}\quad
		\subfigure[McKean, regime I. $\sigma = 0.1, \mu = 1, \lambda = 0$]{\includegraphics[width=.3\textwidth]{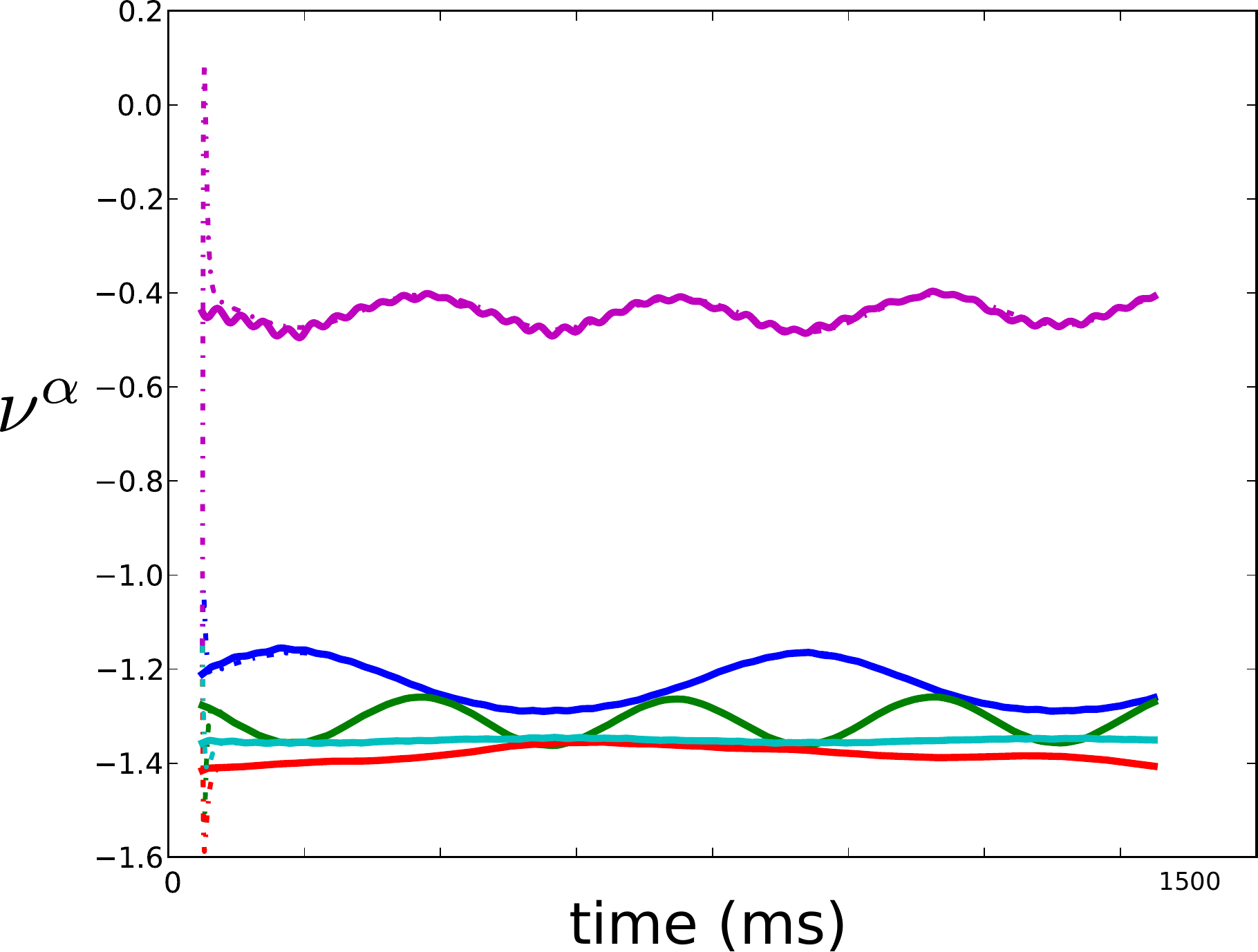}\label{fig: MK comp}}\quad
		\subfigure[Fitzhugh-Nagumo, regime I. $\sigma = 0.5, \mu = 1, \lambda = 0$]{\includegraphics[width=.3\textwidth]{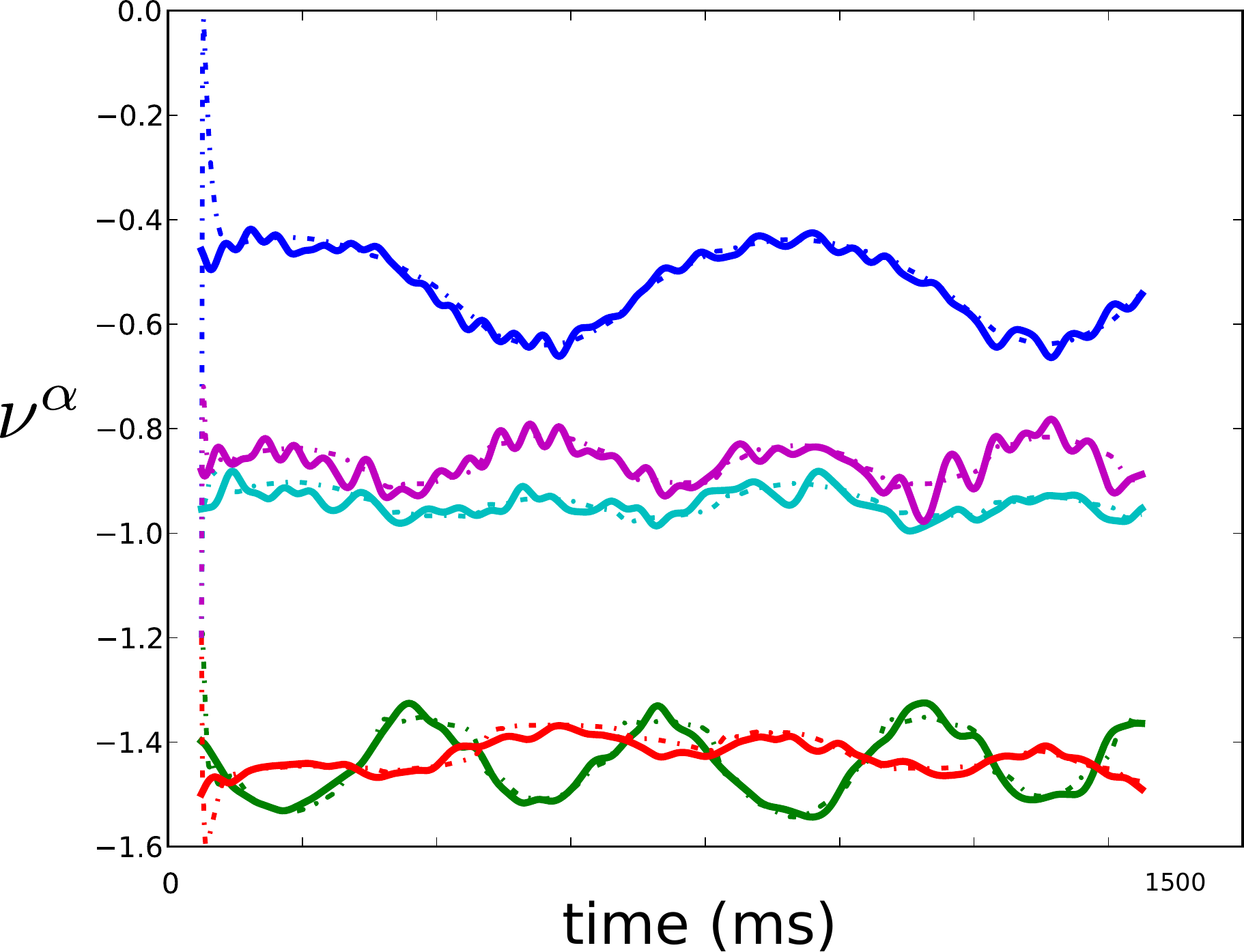}\label{fig: FhN comp}}\\
		\subfigure[Fitzhugh-Nagumo, regime I. $\sigma = 0.5, \mu = 1, \lambda = 1$]{\includegraphics[width=.3\textwidth]{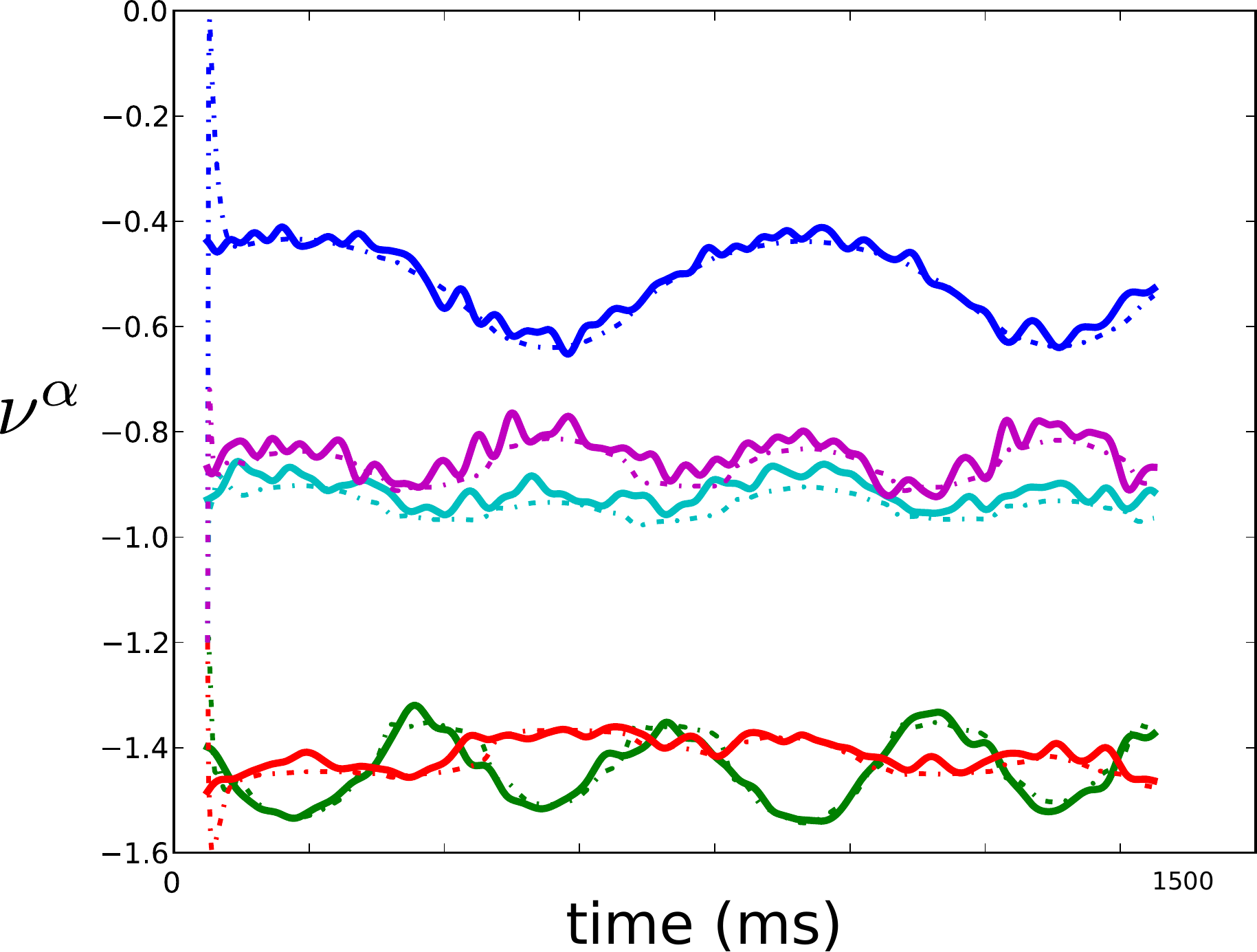}\label{fig: FhN comp lambda}}\quad
		\subfigure[Hodgkin-Huxley, regime I. $\sigma = 1.5, \mu = 0.1, \lambda = 0$]{\includegraphics[width=.3\textwidth]{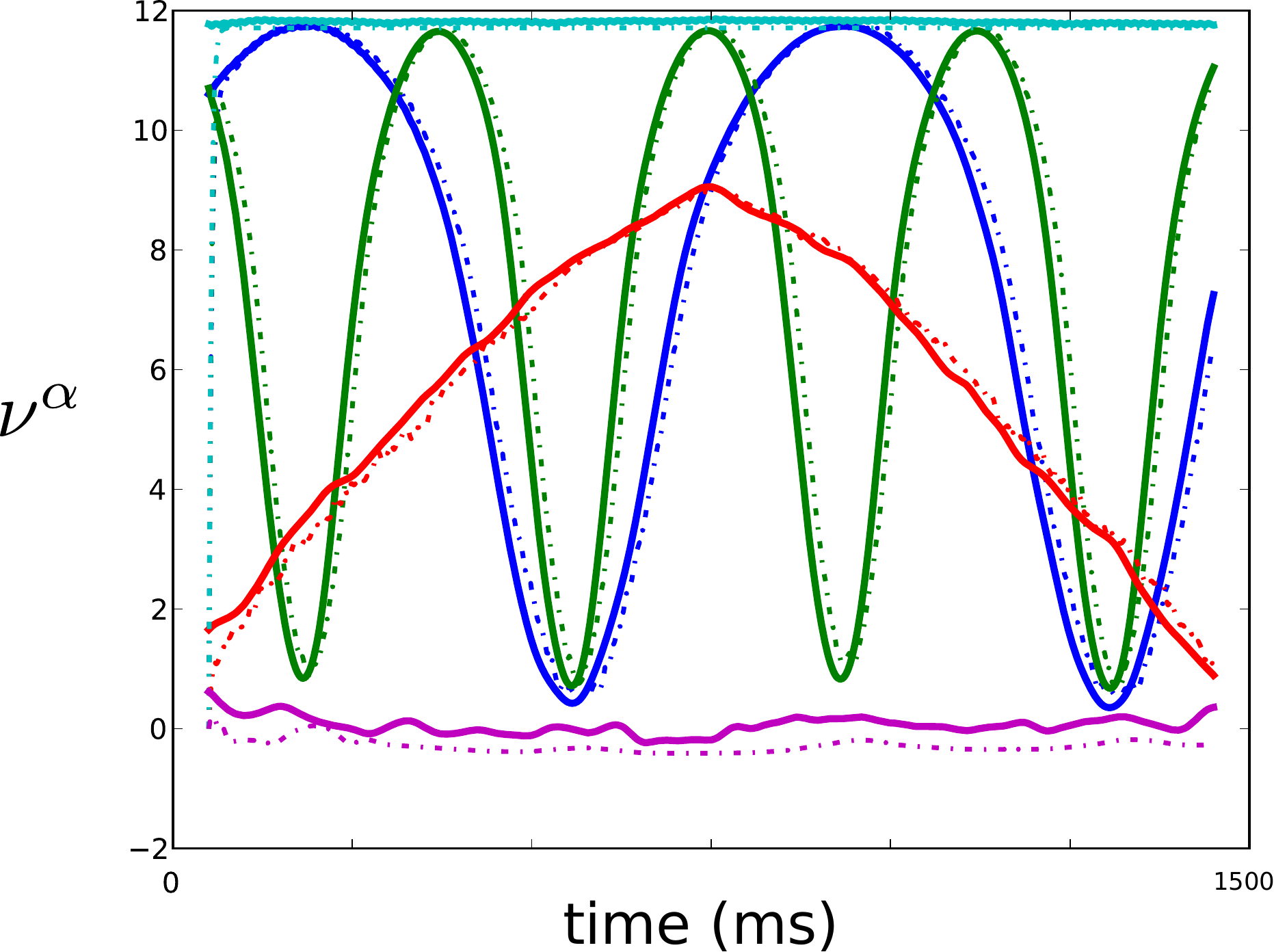}\label{fig: HH comp I}}\quad
		\subfigure[Hodgkin-Huxley, regime II. $\sigma = 0.1, \mu = 0.1, \lambda = 0$]{\includegraphics[width=.3\textwidth]{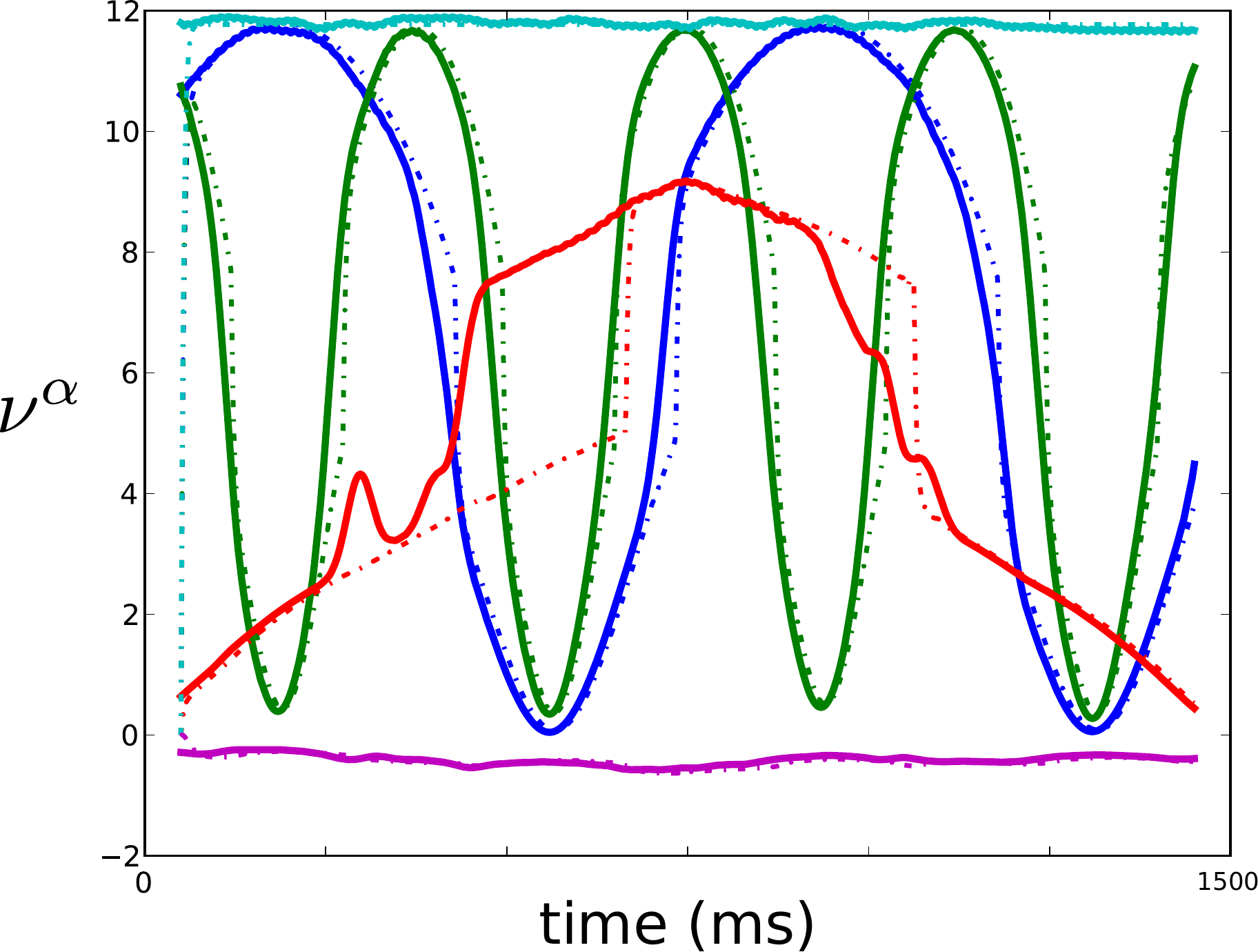}\label{fig: HH comp II}}
		
	\caption{Comparison between the direct computation of the averaged macroscopic variables computed through direct simulation of the network equations (plain lines) and simulations of the macroscopic equations (dashed lines) for networks of $P = 5$ populations with $N_\alpha = 200$ neurons (different populations are depicted in different colors). The connections are drawn randomly according to the Material and Methods description and the random seed is the same for simulations \ref{fig: FhN comp} and \ref{fig: FhN comp lambda}. The inputs $I$ to the McKean and Fitzhugh-Nagumo networks are shown in (a). The inputs to the Hodgkin-Huxley networks correspond to $100I + 10$ where $I$ is shown in (a). The simulations where done using a stochastic Euler algorithm with $T = 15000$ (resp. $T=30000$) timesteps of size $dt=0.1$ (resp. $dt=0.05$) for McKean and FitzHugh-Nagumo (resp. Hodgkin-Huxley) networks.}
	\label{fig:AllComparisons}
\end{figure}

We consider that there are $P=5$ populations of $N=200$ neurons (i.e. $N_\alpha = N_\beta$ for all $\alpha, \beta$). All the neurons in the same population receive the same input corresponding to one color in figure~\ref{fig: inputs}. Some parameters are constant for all simulations:  $\tau_s = 10 ms$, $\theta = 100 ms$. Simulation dependent parameters are detailed in the caption of figure~\ref{fig:AllComparisons} which gathers the comparison for the different models. 

The comparison for the McKean and Fitzhugh-Nagumo neurons show a precise match see figures \ref{fig: MK comp}, \ref{fig: FhN comp}, and \ref{fig: FhN comp lambda} even when the strength $\mu = 1$ of the connections is strong enough to significantly modify the shape of the input signal. Note that, although the mathematical derivation assumed slow inputs and synapses, these simulations validate the reduction even with faster synapses ($\tau_s = 10ms$) and inputs evolving at the scale of hundreds of milliseconds. Also observe that adding frozen noise on the connections with $\lambda = 1$ in figure \ref{fig: FhN comp lambda} does not significantly change the accuracy of the match.

The comparison for the Hodgkin-Huxley neuron are only a relative success. First, the reduced model is only accurate when the connections between populations are kept small, i.e. $\mu = 0.1$ in figure \ref{fig: HH comp I}. This is why the simulations for the Hodgkin-Huxley networks show patterns very similar to the inputs. When the connection strength increases the quality of the match significantly decreases (not displayed here). Second, the algorithm proposed in Material and Methods to simulate the regime II networks fails reproducing faithfully the averaged spiking network, see figure \ref{fig: HH comp II}. More precisely the jump from one branch to the other is not well approximated. We observe that the populations having fast switches in the multivalued region (blue and green) or that do not intersect the multivalued input region (cyan and purple) are precisely recovered. However, the red population, spending much time in the multivalued region, is not well approximated by the macroscopic activity model: 
the network equations may randomly switch from spike to rest, which produces the irregular macroscopic activity, whereas the smooth firing rate does not show such switches. Notwithstanding these unavoidable errors, we observe a fair fit of the macroscopic activity model, which recovers most of the qualitative properties of the network activity.

\section*{Discussion}
Even if collective phenomena arising in large noisy spiking neural networks are extremely complex, we have shown that the macroscopic activity of the network can be consistently described by simple low dimensional deterministic differential equations. The parameters and non-linearity involved are determined by the type of neurons considered. Depending on the neuron model the non-linearity can be a well-behaved function (which we call regime I) or a more complicated multivalued function (which we call regime II in case of two values). The three neuron models we considered (McKean, Fitzhugh-Nagumo and Hodgkin-Huxley) are regime I when the intrinsic noise is strong in the network. However, with weak noise the Hodgkin-Huxley model is regime II, in which case the low-dimensional model proposed is more complicated (involving jumps between attractors). Comparisons of the averaged dynamics of spiking networks with the reduced equations showed a very precise fit, even for initial conditions independent of the network'
s initial conditions, for regime I neuron models. However, for the regime II neuron models, the reduction accuracy is not as good. Indeed, noise will induce random switches from one attractor to the other, which cannot be handled through reduced methods. Yet, the reduced model recovers the main qualitative features of the signal, but in the bistable regions, quantitative distinctions arise. 

The reduction accuracy is significantly better for McKean and Fitzhugh-Nagumo neuron models than for the Hodgkin-Huxley model. Indeed, the reduction for the latter becomes irrelevant for strong connections between neurons whereas it is not the case for the former. We believe this is not due an inherent difference between the models, but rather an incapacity from us to determine a relevant linear part $L$ for the Hodgkin-Huxley model. Indeed, as shown in table \ref{table: linear part}, there is some time-delayed information in the linear part of McKean and Fitzhugh-Nagumo whereas there is simply a linear instantaneous decay for Hodgkin-Huxley model. Further work on reducing Hodgkin-Huxley networks should include a way to add information in the linear part of this model.

This reduction holds on a number of assumptions imposed by the mathematical approach: we need connections within populations to be homogeneous and scaled by the number of neurons in the population, which has to be large enough for averaging effects to occur (i.e. for the mean-field reduction to hold). Besides, the inputs to the neurons in the same population are considered identical. More importantly, the reduction is largely based on the linearity of synapses.  Although the slowness of synapses and inputs were required for the mathematical derivation, simulations have shown that the reduction was quite robust to increased speed for synapse and inputs. It is also fair to observe, that this reduction can not be applied directly to real biological tissues. All the neurons are assumed identical and the propagation of action potential along axons (which is important in measurements) has been ignored. In the future, an extensive work involving detailed biological knowledge and heavy computer simulations could make possible to apply this reduction to biological neural assemblies.

It is important to note the huge complexity reduction obtained: in the case of Hodgkin-Huxley networks, we reduced a system of $4P\,N$ stochastic differential equations with $P$ populations into a system of $P$ deterministic, ordinary one dimensional differential equations. The nonlinear transforms computed, as well as code for the simulations, are provided freely online. For efficient simulation of large-scale neuronal spiking networks with noise, if one is interested in computing the mean macroscopic activity, simulating the reduced model is a precise and simple choice that shall be considered for efficiency. 

This study quantifies the stabilization properties of the noise, that were already discussed in~\cite{wilson-cowan:72}, controlling the shape of the effective non-linearity of the reduced model. Noise tends to act as a linearizer: when the noise is strong, the network macroscopic activity tend to evolve more linearly. The example of Hodgkin-Huxley model shows it can even change a neuron model from regime II to regime I. This implies that knowing the value of the intrinsic noise in biological tissues could be a good indicator to evaluate their level of non-linearity.

  \bibliographystyle{apalike}
  \bibliography{averaging_bib.bib}

\appendix
\section*{Appendix A: Analytical derivations for the deterministic McKean model}
\begin{figure}[!ht]
  \centering
  \includegraphics[width=0.7\textwidth]{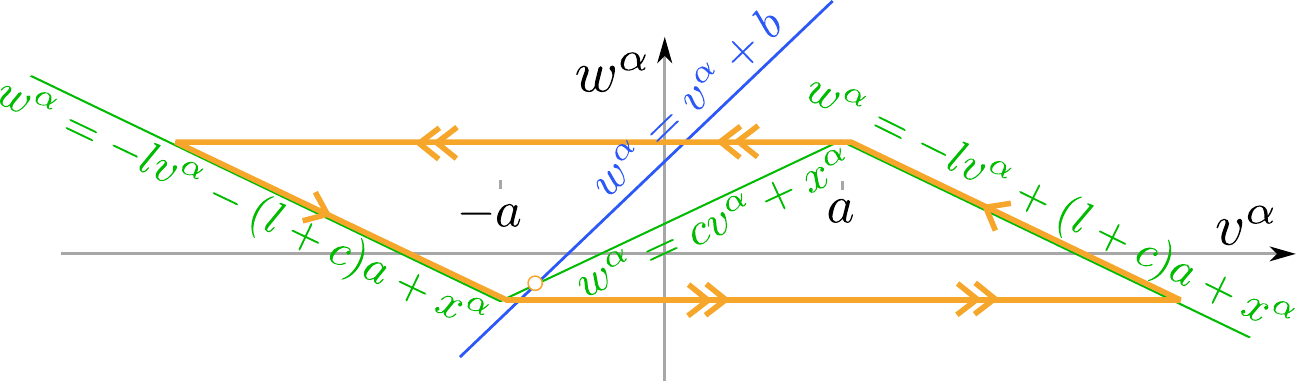}
  \caption{Phase plane of the deterministic McKean neuron (equations~\eqref{eq:MKMFE} with $x^\alpha = \sum_{\beta=1}^p \mathbf{U}^{\alpha\beta}_t + I^\alpha(t)$ a constant). When the blue line and any of the decreasing green line intersect, then there is a stable fixed point. When the blue curve intersect the increasing green line, then there is a periodic orbit (this is the case shown here). The periodic orbit corresponds to the non-smooth orange trajectory composed of two branches on the slow manifold (single-arrowed segments) and horizontal double-arrowed segments correspond to the fast transitions.}
  \label{fig1: McKean}
\end{figure}

Using a McKean network \eqref{eq: MK network} in the mean field reduction \eqref{eq:MFEGeneral} in the case $\sigma^i = 0$ leads to the reduction
\begin{equation}\label{eq:MKMFE}
\begin{array}{rl}
dv^{\alpha} & = \bigg(f(v^{\alpha}) - w^\alpha + \sum_{\gamma=1}^P J_{\alpha\gamma} \Exp{v^{\gamma}}\ast h + I^\alpha(t)\bigg) dt\\
\dot{w}^\alpha & = \eps_w (v^{\alpha} - w^\alpha + b)\\
  \end{array}
\end{equation}
The implicit integration of the adaptation $w^\alpha$ equation in~\eqref{eq:MKMFE} shows that $w^\alpha=v^{\alpha} \ast h_{\tau_w} + b$, where $h_{\tau_w}(t)= \eps_w e^{-\eps_w}\mathbbm{1}_{t>0}$. Using the commutativity of the convolution and equation~\eqref{eq:FiringRateGeneral}, we obtain the exact macroscopic equation for the McKean neuron:
\begin{equation}
\dot{\nu}^\alpha = \E\big(f(v^{\alpha})\big) \ast g - \nu^{\alpha} \ast h_{\tau_w}  - b + \sum_{\beta=1}^P J_{\alpha \beta} \big( \nu^{\beta} \ast h\big) + \tilde{I}^{\alpha}(t)
\label{eq1: evolution firing-rate exact}
\end{equation}
The only unknown term in the formula above is $\E\big(f(v^{\alpha})\big) \ast g$. To compute this term, consider system \eqref{eq:MKMFE} where $x^\alpha = \sum_{\gamma=1}^P J_{\alpha\gamma} \Exp{v^{\gamma}_t}\ast h + I^\alpha(t)$ is assumed to be constant. Indeed, as mentioned before, the following derivation relies on the fact that the inputs and the synapses are so slow that the adiabatic assumption holds (i.e. $v^\alpha$ reaches its equilibrium value immediately). Thus, $v^\alpha$ can be considered uncoupled from the others populations and it has the same dynamics as a single McKean neuron, whose phase plane is shown in figure~\ref{fig1: McKean}.
Under the assumption that the recovery variable is very slow ($\eps_w \ll 1$), the state of neurons in population $\alpha$ is essentially projected on one of the two slow manifold, corresponding in the phase plane Figure~\ref{fig1: McKean} to the single-arrowed orange branches of the $v$-nullcline. Extemely fast switches between these two branches of the slow manifold occur when the trajectories reach an extremity of the manifold. Except during the very fast transitions, it holds that $f(v^{\alpha}) = -l v^{\alpha} \pm (l + c) a + x^{\alpha}$ and hence \[\Exp{f(v^{\alpha})} = -l \Exp{v^{\alpha}} + \E(\pm) (l + c) a + x^{\alpha}\]
with $\E(\pm) = \int_{v^{\alpha} > 0} dv^{\alpha}(t) - \int_{v^{\alpha} < 0} dv^{\alpha}(t) = \Prob(v^{\alpha}>0) - \Prob(v^{\alpha}<0)$. Therefore we have:
\begin{equation*}
\E\big(f(v^{\alpha})\big) \ast g = - l \nu^{\alpha} + (l + c)a \big(\Prob(v^{\alpha}>0) - \Prob(v^{\alpha}<0)\big)\ast g + x^\alpha 
\end{equation*}

\begin{figure}[!ht]
  \centering
  \includegraphics[width=0.5\textwidth]{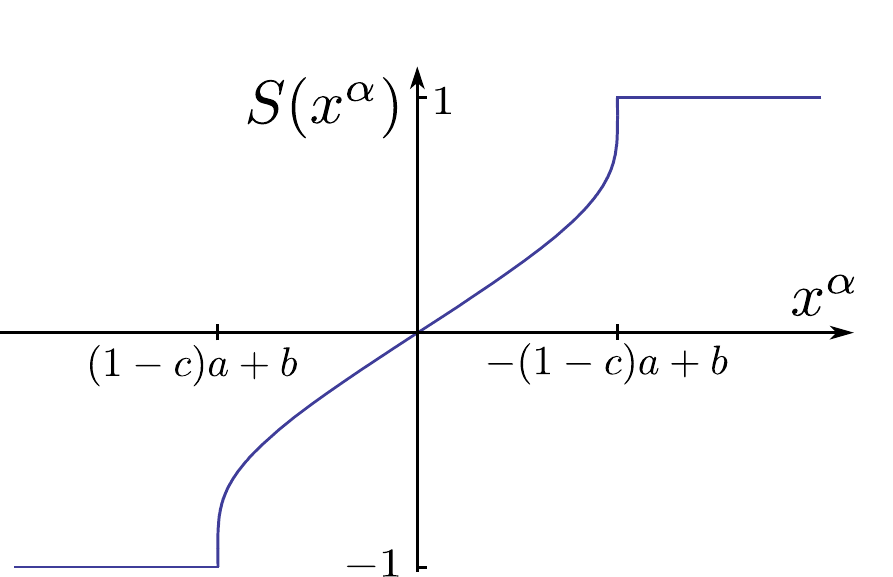}\label{fig1: sigmoid without noise}
  \caption{Function  $S(x^\alpha)$ for the deterministic McKean model given in equation \ref{eq1: sigmoid without noise}.}
  \label{fig: deterministic McKean Sigmoid}
\end{figure}

We write $S\big(x^\alpha\big) \in \R$ the value toward which the function $t \mapsto \Big(\big(\Prob(v^{\alpha}>0) - \Prob(v^{\alpha}<0)\big)\ast g \Big)(t)$ converges when the McKean is stimulated by a constant input $x^\alpha$ (thus we discard the initial transient). Computing $\Prob(v^\alpha \gtrless 0) \ast g$ for a constant effective input $x^\alpha$  amounts to computing the proportion of time system a McKean neuron spends on (or close to) the slow manifolds $w^\alpha = -l v^\alpha \pm (l + c) a + x^\alpha$. This can be performed analytically. Indeed, there are two different cases:
\begin{itemize}
 \item If $x^\alpha \leq -(1-c)a + b$ (resp. $x^\alpha \geq (1-c)a + b$).\\
 Then the system has a single stable fixed point on the negative (resp. positive) slow manifold. In figure \ref{fig1: McKean}, this corresponds to the blue curve crossing the green piecewise cubic where the latter is decreasing. In this case, $S(x^\alpha) = -1$ (resp. $S(x^\alpha) = 1$).
 \item If $-(1-c)a +b < x^\alpha < (1-c)a + b$.\\
 Then the system is oscillating on a deterministic limit cycle represented in orange in figure~\ref{fig1: McKean}. In this case, $S(x^\alpha) = \frac{T^+(x^\alpha) - T^-(x^\alpha)}{T^+(x^\alpha) + T^-(x^\alpha)}$ where $T^-(x^\alpha)$ (resp. $T^+(x^\alpha)$) is the duration it takes for the system to go along the negative (resp. positive) part of the slow manifold. Following \cite{coombes2001phase}, we can access these values. Indeed, assume the fast membrane potential immediately goes to one of the slow nullclines. This gives the equation: $-l v^{\alpha} \pm (l + c)a - w^\alpha + x^\alpha = 0$. Injecting this in the slow equation and integrating along relaxation orbit (orange path in figure \ref{fig1: McKean}) leads to
\begin{equation*}
	T^+(x^\alpha) = \frac{l}{\eps_w}\int_{-c a + x^\alpha}^{c a + x^\alpha} \frac{dw}{-(1 + l)w + (l+c) a + x^\alpha + lb}= \frac{l}{\eps_w(1+l)} \log\Big(\frac{(1 +2c/l + c)a  + b -  x^\alpha}{(1- c)a +b  -  x^\alpha} \Big)
\end{equation*}
Similarly,
$$
  T^-(x^\alpha) = \frac{l}{\eps_w(1+l)} \log\Big(\frac{(1 +2c/l + c)a  +  x^\alpha - b}{(1- c)a  +  x^\alpha - b} \Big)
$$
Therefore, for $x^\alpha \in ]-(1-c)a + b, (1-c)a + b[$
\begin{equation}
  S(x^\alpha) = \frac{\log \Bigg(\frac{\big((1 +2c/l + c)a + b -  x^\alpha \big)\big((1- c)a  +  x^\alpha - b \big)}{\big( (1- c)a + b -  x^\alpha\big)\big((1 +2c/l + c)a  +  x^\alpha - b \big)} \Bigg)}{\log\Bigg(\frac{\big((1 +2c/l + c)a + b -  x^\alpha \big)\big( (1 +2c/l + c)a  +  x^\alpha - b\big)}{\big((1- c)a + b - x^\alpha \big)\big( (1- c)a  +  x^\alpha - b\big)} \Bigg)}
\label{eq1: sigmoid without noise}
\end{equation}
\end{itemize}
This function is shown in figure~\ref{fig: deterministic McKean Sigmoid}. It is a non-smooth sigmoidal function with vertical tangents at $-(1-c)a + b$ and $(1-c)a + b$. This corresponds to the transition from a fixed point to the oscillatory pattern. Note that it is identical to the bifurcation diagram of Fig.~\ref{fig:MKBifs}.

Based on equations \eqref{eq1: evolution firing-rate exact} and the definition of the sigmoid \eqref{eq1: sigmoid without noise}, we are now in position to define an averaged model describing the evolution of the macroscopic population activity. It takes the form of a self consistent, non autonomous, delayed differential system:
\begin{equation}
\dot{\nu}^\alpha =  \underbrace{-\nu^{\alpha} \ast (l \delta + h_{\tau_w})}_{L(\nu^\alpha)} + \tilde{S}\Big(\sum_{\beta=1}^p J_{\alpha \beta} \big(\nu^{\beta} \ast h\big) + I^\alpha \ast g\Big)
\label{eq: MK firing-rate model}
\end{equation}
where $\delta$ is the Dirac function and $\tilde{S}$ is an element-wise function such that $\tilde{S}(x^\alpha) = x^\alpha + (l+c)a S(x^\alpha) - b$.

\end{document}